\documentclass[aps,prd,nofootinbib,twocolumn,superscriptaddress,preprintnumbers,balancelastpage,longbibliography, 10pt]{revtex4-1}
\usepackage{amsmath,amssymb,mathtools,bm}
\usepackage{graphicx, color, hepunits}
\usepackage[dvipsnames]{xcolor}
\usepackage{float}
\usepackage{filecontents}
\usepackage{multirow}

\usepackage[colorlinks=true
,urlcolor=purple
,anchorcolor=blue
,citecolor=red 
,filecolor=magenta
,linkcolor=blue
,menucolor=blue
,linktocpage=true
,pdfproducer=medialab
,pdfa=true
]{hyperref}
\usepackage[utf8]{inputenc}
\usepackage[english]{babel}
\usepackage{soul}
\let\vec\mathbf

\usepackage{overpic}
\usepackage{epsfig}
\usepackage{graphicx}  
\usepackage{url}
\usepackage{color}
\usepackage{hyperref}
\hypersetup{
     colorlinks   = true,
     citecolor    = red,
	 linkcolor=blue
}
\usepackage{float}
\usepackage[export]{adjustbox}
\usepackage{cleveref}

\usepackage[lofdepth,lotdepth,caption=false,justification=RaggedRight]{subfig}

\newcommand{\es}[2] {\begin{equation} \label{#1} \begin{split} #2 \end{split} \end{equation}}

\newcommand{\mpl}{M_{\rm pl}}
\newcommand{\D}{{\rm d}}

\begin{document}

\title{\LARGE{Gravitational Waves from Stochastic Scalar Fluctuations}}

\author{Reza Ebadi}
\affiliation{Department of Physics, University of Maryland, College Park, MD 20742, USA}
\affiliation{Quantum Technology Center, University of Maryland, College Park, MD 20742, USA}

\author{Soubhik Kumar}
\affiliation{Berkeley Center for Theoretical Physics, Department of Physics, University of California, Berkeley, CA 94720, USA}
\affiliation{Theoretical Physics Group, Lawrence Berkeley National Laboratory, Berkeley, CA 94720, USA}

\author{Amara McCune}
\affiliation{Berkeley Center for Theoretical Physics, Department of Physics, University of California, Berkeley, CA 94720, USA}
\affiliation{Theoretical Physics Group, Lawrence Berkeley National Laboratory, Berkeley, CA 94720, USA}
\affiliation{Department of Physics, University of California, Santa Barbara, CA 93106, USA}

\author{Hanwen Tai}
\affiliation{Department of Physics, The University of Chicago, Chicago, IL 60637, USA}

\author{Lian-Tao Wang}
\affiliation{Department of Physics, The University of Chicago, Chicago, IL 60637, USA}
\affiliation{Enrico Fermi Institute, University of Chicago, Chicago, Illinois 60637, USA}
\affiliation{Kavli Institute for Cosmological Physics, University of Chicago, Chicago, Illinois 60637, USA}

\begin{abstract}\noindent
We present a novel mechanism for gravitational wave generation in the early Universe. 
Light spectator scalar fields during inflation can acquire a blue-tilted power spectrum due to stochastic effects. 
We show that this effect can lead to large curvature perturbations at small scales (induced by the spectator field fluctuations) while maintaining the observed, slightly red-tilted curvature perturbations at large cosmological scales (induced by the inflaton fluctuations). 
Along with other observational signatures, such as enhanced dark matter substructure, large curvature perturbations can induce a stochastic gravitational wave background (SGWB).
The predicted strength of SGWB in our scenario, $\Omega_{\rm GW}h^2 \simeq 10^{-20} - 10^{-15}$, can be observed with future detectors, operating between $10^{-5}$~Hz and 10~Hz.
We note that, in order to accommodate the newly reported NANOGrav observation, one could consider the same class of spectator models. At the same time, one would need to go beyond the simple benchmark considered here and consider a regime in which a misalignment contribution is also important.
\end{abstract}

\maketitle
\tableofcontents

\section{Introduction}
The fluctuations observed in the cosmic microwave background (CMB) and large-scale structure (LSS) have given us valuable information about the primordial Universe.
As per the standard $\Lambda$CDM cosmology, such fluctuations were generated during a period of cosmic inflation (see~\cite{Baumann:2009ds} for a review).
While the microphysical nature of inflation is still unknown, well-motivated single-field slow-roll inflationary models predict an approximately scale-invariant spectrum of primordial fluctuations, consistent with CMB and LSS observations.
These observations have enabled precise measurements of the primordial fluctuations between the comoving scales $k\sim 10^{-4}-1~{\rm Mpc}^{-1}$.
However, the properties of primordial density perturbations are comparatively much less constrained for $k\gtrsim {\rm Mpc}^{-1}$.
In particular, as we will discuss below, the primordial curvature power spectrum $\Delta_\zeta^2$ can naturally be much larger at such small scales, compared to the value $\Delta_\zeta^2 \approx 2\times 10^{-9}$ observed on CMB scales~\cite{Planck:2018jri}.

Scales corresponding to $k\gtrsim {\rm Mpc}^{-1}$ are interesting for several reasons.
First, they contain vital information regarding the inflationary dynamics after the CMB-observable modes exit the horizon.
In particular, they can reveal important clues as to how inflation could have ended and the Universe was reheated.
An enhanced power spectrum on such scales can also lead to overabundant dark matter (DM) subhalos, motivating novel probes (see~\cite{Boddy:2022knd} for a review).
Furthermore, if the enhancement is significant, $\Delta_\zeta^2 \gtrsim 10^{-7}$, the primordial curvature fluctuations can induce a stochastic gravitational wave background (SGWB) within the range of future gravitational wave detectors~\cite{Domenech:2021ztg}.
For even larger fluctuations, $\Delta_\zeta^2 \gtrsim 10^{-2}$, primordial black holes (PBH) can form, leading to interesting observational signatures~\cite{Green:2020jor, Carr:2020xqk}.
Given this, it is interesting to look for mechanisms that can naturally lead to a `blue-tilted', enhanced power spectrum at small scales.

In models involving a single dynamical field during inflation, such an enhancement can come, for example, from an inflection point on the inflaton potential or an ultra-slow roll phase~\cite{Ivanov:1994pa, Garcia-Bellido:2017mdw, Ballesteros:2017fsr, Tsamis:2003px, Kinney:2005vj}.\footnote{See also \cite{Hooshangi:2022lao} for PBH formation in a multi-field ultra-slow roll inflationary model.}
However, for any generic structure of the inflaton potential, a power spectrum that is blue-tilted at small scales can naturally arise if there are additional light scalar fields other than the inflaton field.
One class of such mechanisms involves a rolling complex scalar field where the radial mode $\varphi$ has a mass of order the inflationary Hubble scale $H$ and is initially displaced away from the minimum~\cite{Kasuya:2009up}.
As $\varphi$ rolls down the inflationary potential, the fluctuations of the Goldstone mode $\propto (H/\varphi)^2$ increase with time.
This can then give rise to {\it isocurvature} fluctuations that increase with $k$, i.e., a blue-tilted spectrum.
This idea was further discussed in~\cite{Kawasaki:2012wr} to show how {\it curvature} perturbations can be enhanced on small scales as well, and lead to the formation of PBH.
For further studies on blue-tilted isocurvature perturbations, see, e.g.,~\cite{Chung:2015pga, Chung:2017uzc, Chung:2021lfg, Talebian:2022jkb}.
Other than this, models of vector DM~\cite{Graham:2015rva}, early matter domination~\cite{Erickcek:2011us}, and incomplete phase transitions~\cite{Barir:2022kzo} can also give rise to enhanced curvature perturbation at small scales.

In this work, we focus on a different mechanism where a Hubble-mass scalar field quantum mechanically fluctuates around the minimum of its potential, instead of being significantly displaced away from it (as in~\cite{Kasuya:2009up, Kawasaki:2012wr}).\footnote{For scenarios where the spectator field fluctuates around the minimum and gives rise to dark matter abundance, see, e.g.,~\cite{Chung:2004nh}.} 
Hubble-mass fields can naturally roll down to their minimum since the homogeneous field value decreases with time as $\exp(-m^2 t/(3H))$, where $m$ is the mass of the field with $m\lesssim H$.
Given that we do not know the {\it total} number of $e$-foldings that took place during inflation, it is plausible that a Hubble mass particle was already classically driven to the minimum of the potential when the CMB-observable modes exit the horizon during inflation.
For example, for $m^2/H^2 = 0.2$, the field value decreases by approximately a factor of $10^3$, for 100 $e$-foldings of inflation prior to the exit of the CMB-observable modes.
For any initial field value $\varphi_{\rm ini}\lesssim 10^{3} \langle \varphi \rangle$, this can then naturally localize the massive field near the minimum $ \langle \varphi \rangle$.
However, the field can still have quantum mechanical fluctuations which tend to diffuse the field away from $ \langle \varphi \rangle$.
The potential for the field, on the other hand, tries to push the field back to $ \langle \varphi \rangle$.
The combination of these two effects gives rise to a non-trivial probability distribution for the field, both as a function of time and space.

\begin{figure}
    \centering
    \includegraphics[width=0.45\textwidth]{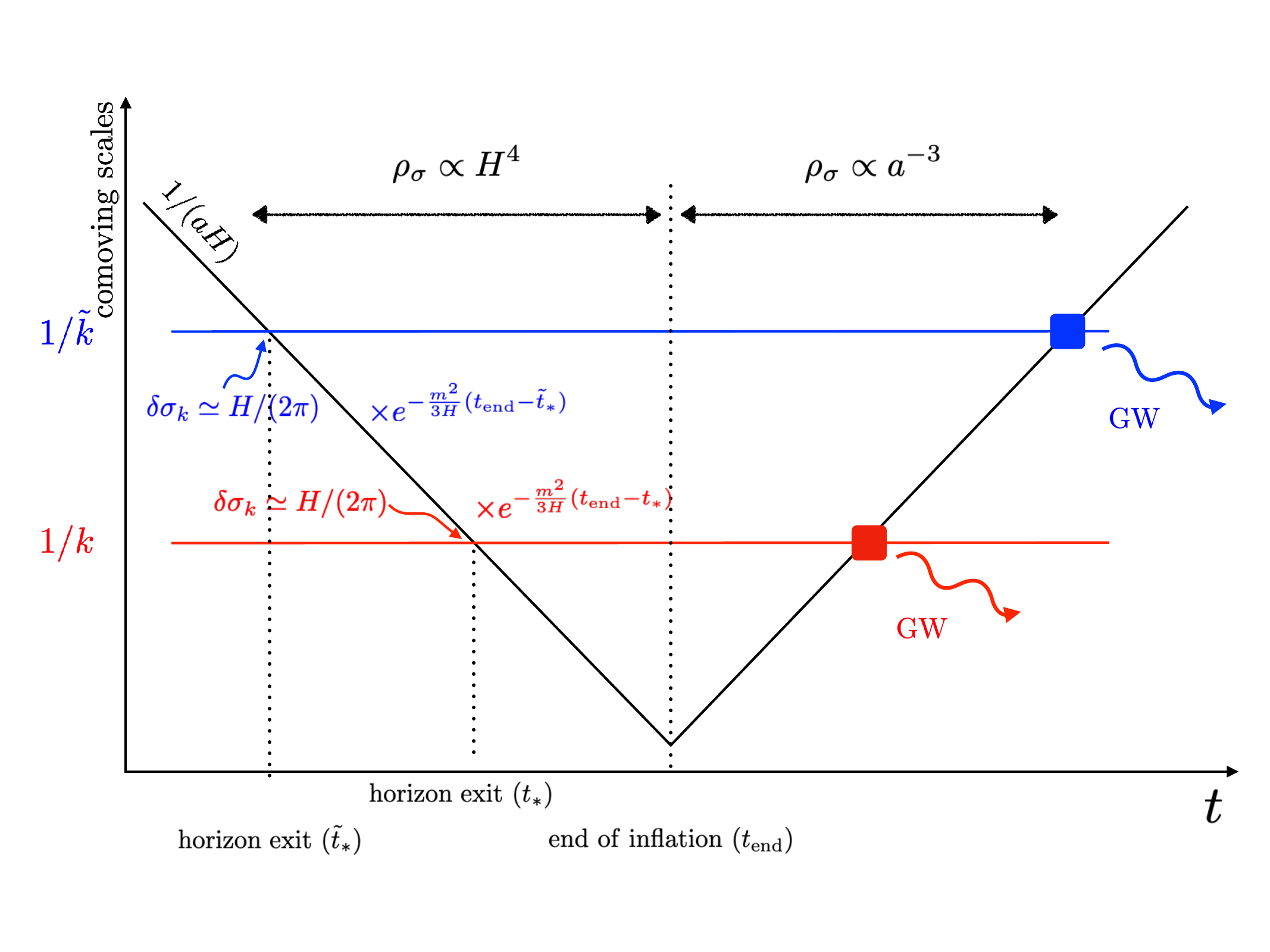}
    \caption{Schematic of the mechanism. The comoving horizon $1/(aH)$ decreases during inflation and increases after that. Any $k$-mode carries a fluctuation of order $H/(2\pi)$ at the time of mode exit. However, modes with larger $k$ (red) exit the horizon later and encounters less dilution compared to modes with smaller $k$ (blue), since $t_*>\tilde{t}_*$. Consequently, modes with larger $k$ source stronger gravitational waves upon horizon re-entry (shown via square box). We also depict the fact that $\sigma$ carries an energy density $\propto H^4$ during inflation, and dilutes as matter (for our benchmark choices) after inflation ends.}
    \label{fig:schematic}
\end{figure}

We study these effects using the stochastic formalism~\cite{Starobinsky:1986fx, Starobinsky:1994bd} for light scalar fields in de Sitter (dS) spacetime.
In particular, such stochastic effects can lead to a spectrum that is blue-tilted at small scales.
While we carry out the computation by solving the associated Fokker-Planck equation in detail below, we can intuitively understand the origin of a blue-tilted spectrum as follows.
For simplicity, we momentarily restrict our discussion to a free scalar field $\sigma$ with mass $m$ such that $m^2\lesssim H^2$.
The fluctuation $\sigma_k(t) $, corresponding to a comoving $k$-mode, decays after horizon exit as $\sigma_k(t) \sim H \exp(-m^2(t-t_*)/(3H))$, where $t_*$ is the time when the mode exits the horizon, $k = a(t_*) H$.
We can rewrite the above by noting that physical momenta redshift as a function of time via $k/a(t) = H \exp(-H(t-t_*))$.
Then we arrive at, $\sigma_k(t) \sim H (k/(aH))^{m^2/(3H^2)}$.
Therefore, the dimensionless power spectrum, $|\sigma_k|^2 k^3 \propto (k/(aH))^{2m^2/(3H^2)}$ has a blue tilt of $2m^2/(3H^2)$.
Physically, modes with smaller values of $k$ exit the horizon earlier and get more diluted compared to modes with larger values of $k$, leading to more power at larger $k$, and thus a blue-tilted spectrum.
This qualitative feature, including the specific value of the tilt for a free field, is reproduced by the calculation described later where we also include the effects of a quartic self-coupling.
We summarize the mechanism in Fig.~\ref{fig:schematic}.

We note that if $m$ is significantly smaller than $H$, the tilt is reduced and the observational signatures are less striking.
On the other hand, for $m\gtrsim H$, the field is exponentially damped, and stochastic effects are not efficient in displacing the field away from the minimum.
Therefore, it is puzzling as to why the particle mass, a priori arbitrary, could be close to $H$ in realistic scenarios.
However, a situation with $m\approx H$ can naturally rise if the field is non-minimally coupled to gravity.
That is, a coupling ${\cal L}\supset c R \sigma^2$, where $R$ is the Ricci scalar, can uplift the particle mass during inflation $m^2   = (c/12) H^2$, regardless of a smaller `bare' mass.
Here we have used $R=(1/12)H^2$ during inflation, and we notice for $c\sim {\cal O}(1)$, we can have a non-negligible blue-tilted spectrum. 

The way the spectrum of $\sigma$ affects the curvature perturbation depends on the cosmology, and in particular, the lifetime of $\sigma$.
During inflation, the energy density stored in $\sigma$ is of order $H^4$, as expected, since $\sigma$ receives $H$-scale quantum fluctuations.
This is subdominant compared to the energy stored in the inflaton field $\sim H^2\mpl^2$.
This implies $\sigma$ acts as a spectator field during inflation, and through the stochastic effects, $\sigma$ obtains isocurvature fluctuations.
After the end of inflation, $\sigma$ dilutes as matter while the inflaton decay products dilute as radiation.
Therefore, similar to the curvaton paradigm~\cite{Linde:1996gt, Enqvist:2001zp, Moroi:2001ct, Lyth:2001nq}, the fractional energy density in $\sigma$ increases with time.
Eventually, $\sigma$ decays into Standard Model radiation, and its isocurvature perturbations get imprinted onto the curvature perturbation.
Different from the curvaton paradigm, in our scenario, $\sigma$ does not dominate the energy density of the Universe, and also the fluctuations of the inflaton are not negligible.
In particular, on large scales, observed via CMB and LSS, the fluctuations are red-tilted and sourced by the inflaton, as in $\Lambda$CDM cosmology.
On the other hand, the blue-tilted $\sigma$ fluctuations are subdominant on those scales, while dominant at smaller scales $\lesssim {\rm Mpc}$.
These enhanced perturbations can source an SGWB, observable in future gravitational wave detectors, as we describe below.

The rest of the work is organized as follows.
In \cref{sec:cosmo}, we describe the evolution of the inflaton field and $\sigma$ along with some general properties of curvature perturbation in our framework.
In \cref{sec:stochastic_formalism}, we compute the stochastic contributions to $\sigma$ fluctuations to obtain its power spectrum.
We then use these results in \cref{sec:P_zeta} to determine the full shape of the curvature power spectrum, both on large and small scales.
The small-scale enhancement of the curvature power spectrum leads to an observable SGWB and we evaluate the detection prospects in \cref{sec:GW} in the context of $\mu$-Hz to Hz-scale gravitational wave detectors.
We conclude in \cref{sec:concl}.
We include some technical details relevant to the computation of SGWB in \cref{app:GW_details}.

\section{Cosmological History and Curvature Perturbation}
\label{sec:cosmo}
We now describe in detail the cosmological evolution considered in this work.
We assume that the inflaton field $\phi$ drives the expansion of the Universe during inflation and the quantum fluctuations of $\phi$ generate the density fluctuations that we observe in the CMB and LSS, as in standard cosmology.
We also assume that there is a second real scalar field $\sigma$ which behaves as a subdominant spectator field during inflation, as alluded to above.
We parametrize its potential as,
\es{eq:V_quartic}{
V(\sigma) = {1 \over 2} m^2 \sigma^2 + {1 \over 4} \lambda \sigma^4.
}
The $\sigma$ field does not drive inflation but nonetheless obtains quantum fluctuations during inflation. 
In particular, $\sigma$ obtains stochastic fluctuations around the minimum of its potential, as we compute in \cref{sec:stochastic_formalism}.
After the end of inflation, the inflaton is assumed to reheat into radiation with energy density $\rho_r$, which dominates the expansion of the Universe.

\begin{figure}
    \centering
    \includegraphics[width=0.45\textwidth]{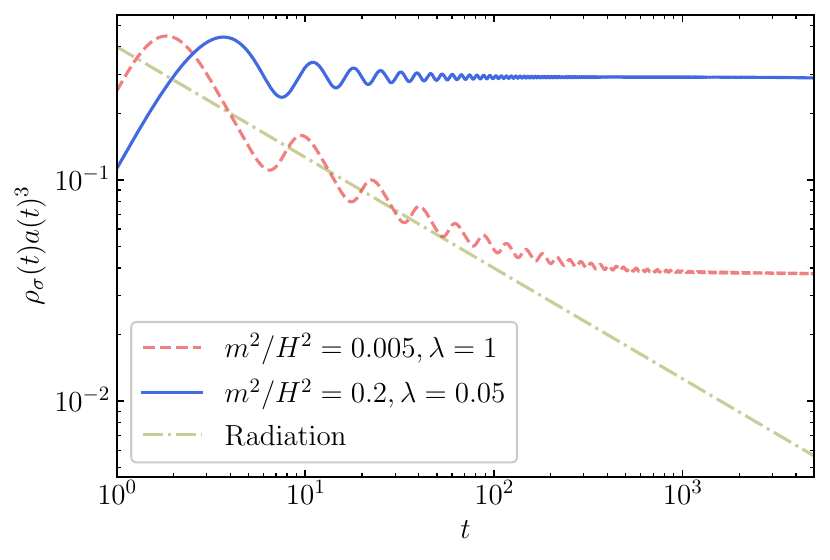}
    \caption{Time evolution of scalar field energy density $\rho_\sigma(t)$. In scenarios where the quartic term dominates the initial evolution (dashed red), the field dilutes as radiation (dot-dashed olive), $\rho_\sigma(t) \propto 1/a(t)^4$. Eventually, the mass term becomes important, and the behavior becomes $\rho_\sigma(t) \propto 1/a(t)^3$. The benchmark choices in this work will mimic the blue curve where the evolution of $\rho_\sigma(t)$ is always dominated by the mass term with a matter-like dilution. For both the blue and the red curves, $t=1$ corresponds to the moment when the Hubble scale is approximately equal to the effective mass and the field starts oscillating.}
    \label{fig:dilution}
\end{figure}

On the other hand, the evolution of the $\sigma$ field depends on its mass $m$, interaction $\lambda$, and its frozen (root mean squared) displacement $\sigma_0$ during inflation.
As long as the `effective' mass of $\sigma$: $m^2 + 3\lambda \sigma_0^2$, is smaller than the Hubble scale, $\sigma$ remains approximately frozen at $\sigma_0$.
However, after the Hubble scale falls below the effective mass, $\sigma$ starts oscillating around its potential.
The evolution of its energy density $\rho_\sigma$, during this oscillatory phase depends on the values of $m$ and $\lambda$.
If the quartic interactions dominate, with $\lambda\sigma^2\gg m^2$, $\rho_\sigma$ dilutes like radiation~\cite{Kolb:1990vq}.
Eventually, the amplitude of $\sigma$ decreases sufficiently, so that $\lambda\sigma^2\lesssim m^2$, following which $\rho_\sigma$ starts redshifting like matter.
We illustrate these behaviors in Fig.~\ref{fig:dilution}.

Similar to the curvaton paradigm~\cite{Linde:1996gt, Enqvist:2001zp, Moroi:2001ct, Lyth:2001nq}, during the epoch $\rho_\sigma$ is diluting as matter, its fractional energy density, $f_\sigma(t) \equiv \rho_\sigma(t)/\rho_r(t)$, increases linearly with the scale factor $a(t)$.
For our benchmark parameter choices, we assume $\sigma$ to decay into SM radiation while $f_\sigma(t_d)\sim 1$, where $t_d$ denotes the time of $\sigma$ decay.
After $t_d$, the evolution of the Universe coincides with standard cosmology.
 
With this cosmology in mind, we can track the evolution of various cosmological perturbations using the gauge invariant quantity $\zeta$, the curvature perturbation on uniform-density hypersufaces~\cite{Malik:2008im},
\es{eq:zeta}{
\zeta = -\psi-H{\delta\rho \over \dot{\rho}}.
}
Here $\psi$ is a fluctuation appearing in the spatial part of the metric as, $\delta g_{ij}= -2a^2\psi \delta_{ij}$ (ignoring vector and tensor perturbations), $\delta\rho$ denotes a fluctuation around a homogeneous density $\rho$, and an overdot denotes a derivative with respect to physical time $t$. 
We assume that the decay products of $\phi$ do not interact with $\sigma$ during their cosmological evolution.
Since there is no energy transfer between the two sectors, their energy densities evolve as,
\begin{equation}
    \dot{\rho}_r = -4H\rho_r\,, \quad \dot{\rho}_\sigma = -3H\rho_\sigma,
\end{equation}
where we have focused on the epoch where $\sigma$ dilutes like matter.
For the benchmark parameter choices discussed below, the matter-like dilution for $\sigma$ onsets soon after inflation.
Similar to~\cref{eq:zeta}, we can parametrize gauge invariant fluctuations in radiation and $\sigma$ with the variables, 
\es{}{
\zeta_r = -\psi + \dfrac{1}{4}\dfrac{\delta\rho_r}{\rho_r},~\zeta_\sigma = -\psi + \dfrac{1}{3}\dfrac{\delta\rho_\sigma}{\rho_\sigma}.
}
In terms of the above variables, we can express \cref{eq:zeta} as,
\begin{equation}
    \zeta = {4 \over {4+3f_\sigma}}\zeta_r + {3 f_\sigma \over {4+3f_\sigma}} \zeta_\sigma = \zeta_r + {f_\sigma \over {4+3 f_\sigma}} S_\sigma.
\end{equation}
Here $S_\sigma \equiv 3(\zeta_\sigma - \zeta_r)$ is the isocurvature perturbation between radiation and $\sigma$ perturbations.
In the absence of any energy transfer, $\zeta_r$ and $\zeta_\sigma$ are each conserved at super-horizon scales~\cite{Wands:2000dp}. 
As a result, the evolution of $\zeta$ is entirely determined by the time-dependent relative energy density of between radiation and $\sigma$, $f_\sigma = \rho_\sigma/\rho_r$.
Since $\zeta_r$ and $S_\sigma$ are uncorrelated, the power spectrum for curvature perturbation $ \langle \zeta(\vec{k})\zeta(\vec{k')\rangle}\equiv (2\pi)^3\delta(\vec{k}+\vec{k'})P_\zeta(k)$ is determined by,
\begin{equation}\label{eq:final_Pzeta}
    P_\zeta(k) = P_{\zeta_r}(k) + \left(\frac{f_\sigma}{4+3f_\sigma}\right)^2 P_{S_\sigma}(k)\,,
\end{equation}
or equivalently,
\begin{equation}\label{eq:Delta_zeta}
    \Delta^2_\zeta(k) = \Delta^2_{\zeta_r}(k) + \left(\frac{f_\sigma}{4+3f_\sigma}\right)^2 \Delta^2_{S_\sigma}(k)\,,
\end{equation}
where $\Delta_\zeta^2(k) = k^3P_\zeta(k)/(2\pi^2)$, with $\Delta_{\zeta_r}^2(k)$ and $\Delta_{S_\sigma}^2(k)$ defined analogously.

To compute the spectral tilt, we denote the comoving momentum of the mode that enters the horizon at $t_d$, the time of $\sigma$ decay, as $k_d$ which satisfies $k_d = a(t_d) H(t_d)$.
For $t>t_d$, $\zeta$ remains conserved with time on superhorizon scales.
Correspondingly, for $k<k_d$, the spectral tilt is given by,
\es{eq:tilt_small_k}{
n_s - 1 \equiv &~ {{\D\ln\Delta_\zeta^2(k)} \over {\D\ln k}} = {\Delta_{\zeta_r}^2(k) \over \Delta_\zeta^2(k)}{{\D\ln\Delta_{\zeta_r}^2(k)} \over {\D\ln k}}\\
+&~\left(\frac{f_\sigma}{4+3f_\sigma}\right)^2{\Delta_{S_\sigma}^2(k) \over \Delta_\zeta^2(k)}{{\D\ln\Delta_{S_\sigma}^2(k)} \over {\D\ln k}}.
}
We will consider scenarios where the radiation energy density $\rho_r$ originates from the inflaton, and therefore, $\D\ln\Delta_{\zeta_r}^2(k) / \D\ln k \approx -0.04$ determines the spectral tilt observed on CMB scales~\cite{Planck:2018jri}.
On the other hand, $\sigma$ acquires stochastic fluctuations to give rise to a blue-tilted power spectrum with $\D\ln\Delta_{S_\sigma}^2(k) / \D\ln k \sim 0.3$, as discussed next in \cref{sec:stochastic_formalism}.
Since we will be interested in scenarios with $f_\sigma \lesssim 1$, i.e., $(f_\sigma/(4+3f_\sigma))^2 \lesssim 0.02$, we require $\Delta_{S_\sigma}^2(k) / \Delta_\zeta^2(k) \lesssim 1$ on CMB-scales to be compatible with CMB measurements of $n_s$.
We can also compute the running of the tilt,
\es{eq:running_tilt}{
{\D n_s \over \D\ln k} \approx  \left(\frac{f_\sigma}{4+3f_\sigma}\right)^2 {\Delta_{S_\sigma}^2(k) \over \Delta_\zeta^2(k)}{\left({\D\ln\Delta_{S_\sigma}^2(k)} \over {\D\ln k}\right)^2}.
} 
Our benchmark parameter choices, discussed above, thus also satisfy the CMB constraints on $\D n_s/ \D\ln k$~\cite{Planck:2018jri}.

\section{Review of the Stochastic Formalism}
\label{sec:stochastic_formalism}
A perturbative treatment of self-interacting light scalar fields in de Sitter (dS) spacetime is subtle due to infrared divergences. 
A stochastic approach~\cite{Starobinsky:1986fx, Starobinsky:1994bd} can be used to capture the nontrivial behavior of such fields in dS. 
In this formalism, the super-horizon components of the fields are considered classical stochastic fields that satisfy a Langevin equation, which includes a random noise originating from the sub-horizon physics. 
This gives rise to a Fokker-Planck equation for the probability distribution function (PDF) of the stochastic field, which can be used to calculate correlation functions of physical observables.
We now review these ideas briefly while referring the reader to refs.~\cite{Starobinsky:1986fx, Starobinsky:1994bd, Sasaki:1987gy, Nambu:1987ef, Graham:2018jyp, Markkanen:2019kpv} for more details. 

\subsection{Langevin and Fokker-Planck Equations}

The stochastic approach provides an effective description for the long-wavelength, superhorizon sector of the field theory by decomposing the fields into long-wavelength classical components and short-wavelength quantum operators. 
For instance, a light scalar field can be decomposed as
\begin{equation}
\begin{aligned}
	&\sigma_{\mathrm{tot.}}(\mathbf{x},t) = \sigma(\mathbf{x},t) \\&+ \int\dfrac{\mathrm{d}^3k}{(2\pi)^3}\theta(k-\epsilon a(t)H)e^{-i\mathbf{k}\cdot\mathbf{x}}(a_{\mathbf{k}}u_{k}+a^\dagger_{-\mathbf{k}}u^\ast_{k}),
\end{aligned}
\end{equation}
where $\theta(\cdots)$ is the Heaviside step function, $a$ is the scale factor, $H$ is the Hubble scale, and $\epsilon\lesssim 1$ is a constant number (not to be confused with the slow-roll parameter) which defines the boundary between long ($k<\epsilon a(t) H$) and short ($k>\epsilon a(t) H$) modes. 
We have also denoted the classical part of the field as $\sigma(\mathbf{x},t)$.
The quantum description of the short modes is characterized by the creation and annihilation operators $a_{\mathbf{k}},a^\dagger_{\mathbf{k}}$ along with the mode functions $u_k(t),u_k^*(t)$.

For a light field with $|V''(\sigma)|\ll H^2$, it can be shown~\cite{Starobinsky:1986fx, Starobinsky:1994bd, Sasaki:1987gy, Nambu:1987ef} that the classical part of the field, $\sigma(\mathbf{x},t)$, follows a Langevin equation
\begin{equation}
\label{eq.langevin}
    \dot{\sigma}(\mathbf{x},t) = -\dfrac{1}{3H}V'(\sigma) + \xi(\mathbf{x},t).
\end{equation}
Here an overdot and a prime denote derivative with respect to time and the field, respectively.
The noise $\xi$ arises from short-scale modes,
\begin{equation}
    \xi(\mathbf{x},t) = \epsilon a H^2\int\dfrac{\mathrm{d}^3k}{(2\pi)^3}\delta(k-\epsilon aH)e^{-i\mathbf{k}\cdot\mathbf{x}}(a_{\mathbf{k}}u_{k}+a^\dagger_{-\mathbf{k}}u^\ast_{k}),
\end{equation}
with a correlation
\begin{equation}
    \langle\xi(\mathbf{x}_1,t_1)\xi(\mathbf{x}_2,t_2)\rangle = \dfrac{H^3}{4\pi^2}\delta(t_1-t_2)j_0(\epsilon a H|\mathbf{x}_1-\mathbf{x}_2|),
\end{equation}
where $j_0(x)=\sin x/x$ is the zeroth order spherical Bessel function. 
We see that the noise is uncorrelated in time (i.e., it is a white noise), but also it is uncorrelated over spatial separations larger than $(\epsilon a H)^{-1}$.

The Langevin equation~\eqref{eq.langevin} gives rise to a Fokker-Planck equation for the one-point PDF,
\es{eq:Fokker_Planck}{
&\dfrac{\partial P_{\text{\tiny FP}}(t,\sigma(\mathbf{x},t))}{\partial t} = \left[\dfrac{V''(\sigma(\mathbf{x},t))}{3H}\right.\\
&+\left.\dfrac{V'(\sigma(\mathbf{x},t))}{3H}\dfrac{\partial}{\partial \sigma} +\dfrac{H^3}{8\pi^2}\dfrac{\partial^2}{\partial\sigma^2}\right]P_{\text{\tiny FP}}(t,\sigma(\mathbf{x},t)).
}
Here $P_{\text{\tiny FP}}(t,\sigma(\mathbf{x},t))$ is the PDF of the classical component to take the value $\sigma(\mathbf{x},t)$ at time $t$. 
Thus the Fokker-Planck equation describes how an ensemble of field configurations evolves as a function of time, according to the underlying Langevin equation. 
In this equation, the first and second terms on the right-hand side represent classical drift terms that depend on the potential $V(\sigma)$.
The third term represents a diffusion contribution from the noise $\xi$.
While the classical drift tries to move the central value of the field towards the minimum of the potential, the diffusion contribution pushes the field away from the minimum.
An equilibrium is achieved when these two effects balance each other.
This equilibrium solution can be obtained by setting $\partial P_{\text{\tiny FP}}/\partial t=0$ in~\eqref{eq:Fokker_Planck}, and is given by
\begin{equation}\label{eq:P_eq}
    P_{\text{\tiny FP},\mathrm{eq}}(\sigma) = \dfrac{1}{\mathcal{N}}\exp\left(-\dfrac{8\pi^2}{3H^4}V(\sigma)\right),
\end{equation}
where $\mathcal{N}$ is a normalization constant.
Upon a variable change 
\begin{align}
	\tilde{P}_{\text{\tiny FP}}(t,\sigma)\equiv \exp\left(\frac{4\pi^2V(\sigma)}{3H^4}\right)P_{\text{\tiny FP}}(t,\sigma),
\end{align}
eq.~\eqref{eq:Fokker_Planck} can written as
\begin{align}\label{eq:Ptilde_FP_eq}
    \dfrac{\partial \tilde{P}_{\text{\tiny FP}}(t,\sigma)}{\partial t} &= \dfrac{H^3}{4\pi^2}\underbrace{\left[-\dfrac{1}{2}\left(v'^{2}-v''\right) +\dfrac{1}{2}\dfrac{\partial^2}{\partial\sigma^2}\right]}_{D_\sigma}\tilde{P}_{\text{\tiny FP}}(t,\sigma)\,,\end{align}
with $v(\sigma) = 4\pi^2 V(\sigma)/(3H^4)$.
We can recast the above as an eigenvalue equation. 
To that end, we write
\begin{align}
	\tilde{P}_{\text{\tiny FP}}(t,\sigma) = \sum_n a_n e^{-\Lambda_n t} \psi_n(\sigma),
\end{align}
where $\psi_n(\sigma)$ satisfies the equation
\begin{equation}\label{eq:eigenvalue_eq}
    D_\sigma\psi_n(\sigma) = -\dfrac{4\pi^2}{H^3}\Lambda_n\psi_n(\sigma).
\end{equation}
The eigenfunctions $\psi_n(\sigma)$ form an orthonormal basis of functions and $a_n$'s are some arbitrary coefficients.

This time-independent eigenvalue equation~\eqref{eq:eigenvalue_eq} can be solved numerically for a generic potential $V(\sigma)$, as we discuss below with an example.
By definition, and independent of the form of the potential, the eigenfunction $\psi_0$ corresponding to the eigenvalue $\Lambda_0=0$, determines the equilibrium distribution. 
Solution of the eq.~\eqref{eq:eigenvalue_eq} for $\Lambda_0=0$ is given by
\begin{equation}
\label{eq.ground}
    \psi_0(\sigma) = \frac{1}{{\sqrt{\cal N}}} \exp\left(-\dfrac{4\pi^2}{3H^4} V(\sigma)\right)\,.
\end{equation}
Thus comparing to eq.~\eqref{eq:P_eq} we get,
\begin{equation}
    P_{\text{\tiny FP},\mathrm{eq}}(\sigma) = \psi_0(\sigma)^2\,.
\end{equation}

\subsection{Two-point Correlation Function and Power Spectrum}
We are interested in calculating the two-point correlation functions of cosmological perturbations.
Any such two-point correlation function depends only on the geodesic distance $s$ between the two points.
Given the coordinates of the two points $(\mathbf{x}_1, t_1)$ and $(\mathbf{x}_2, t_2)$, this distance can be parametrized by $z = 1 + H^2 s^2/2$ with
\begin{equation}\label{eq:dS_invariant_y}
    z = \cosh H(t_1-t_2) - \dfrac{1}{2}e^{H(t_1+t_2)}\left(H|{\mathbf{x}_1-\mathbf{x}_2}|\right)^2.
\end{equation}
To understand the significance of the variable $z$, we first write the two-point correlation function for an arbitrary function of $\sigma$, $g(\sigma)$, as
\begin{equation}
    G_g(\mathbf{x}_1,t_1;\mathbf{x}_2,t_2) = \langle g(\sigma(\mathbf{x}_1,t_1)) g(\sigma(\mathbf{x}_2,t_2)) \rangle.
\end{equation}
To compute this, it is more convenient to calculate the temporal correlation first, and then use the fact that equal-time correlations over spatially separated points are related to the temporal correlation through the de Sitter-invariant variable $z$~\eqref{eq:dS_invariant_y}.
In particular, for coincident points $G_g$ is a function of $(t_1-t_2)$ only, which can be expressed in terms of $z$ for large $|z|$ as,
\begin{align}\label{eq:temporal_corr}
	G_g(t_1 - t_2) \approx G_g(H^{-1}\ln|2z|).
\end{align} 
However, for an equal time correlation function we can also write,
\begin{align}
	|2z| \approx (H e^{Ht}|\mathbf{x}_1-\mathbf{x}_2|)^2,
\end{align} 
which gives,  
\begin{equation}\label{eq:time_corr_to_spatial_corr}
    G_g(t_1-t_2) \simeq G_g\left({\ln|2z|\over H}\right) \simeq G_g\left(\frac{2}{H}\ln(aH|{\mathbf{x}_1-\mathbf{x}_2}|)\right),
\end{equation}
where the approximations hold as long as $|z|\gg1$ and we used $a(t) = \exp(Ht)$.

Now we aim at formally calculating $G_g(t)$ in terms of solutions of the Fokker-Planck equation. The temporal correlation can be written as (see, e.g.,~\cite{Starobinsky:1986fx, Starobinsky:1994bd, Markkanen:2019kpv})
\begin{equation}\label{eq:G_g_init}
    G_g(t) = \int\mathrm{d}\sigma\int\mathrm{d}\sigma_0 P_{\text{\tiny FP},\mathrm{eq}}(\sigma_0) g(\sigma_0) \Pi(t,\sigma;\sigma_0) g(\sigma),
\end{equation}
where $\Pi(t,\sigma;\sigma_0)$ is the kernel function of the time evolution of the probability distribution function, i.e., if the probability distribution is $\delta(\sigma-\sigma_0)$ at $t=0$ it would be $\Pi(t,\sigma;\sigma_0)$ at time $t$. 
In particular, it is defined by
\begin{equation}
    P_{\text{\tiny FP}}(t; \sigma) = \int\mathrm{d}\sigma_0\Pi(t,\sigma;\sigma_0) P(0;\sigma_0).
\end{equation}
In terms of re-scaled probabilities, we can rewrite the above as,
\begin{align}
    \tilde{P}_{\text{\tiny FP}}(t; \sigma) &= \int\mathrm{d}\sigma_0\tilde{\Pi}(t,\sigma;\sigma_0) \tilde{P}_{\text{\tiny FP}}(0;\sigma_0)\,,\\\quad \Pi(t,\sigma;\sigma_0) &= e^{-v(\sigma)} \tilde{\Pi} (t,\sigma;\sigma_0)e^{v(\sigma_0)}.
\end{align}
It follows that $\tilde{\Pi}$ satisfies the same Fokker-Planck equation as $\tilde{P}_{\rm FP}$ \eqref{eq:Ptilde_FP_eq}. 
Therefore, the solutions can be written as
\begin{equation}
    \tilde{\Pi}(t;\sigma,\sigma_0) = \sum_n \psi_n(\sigma_0) e^{-\Lambda_nt}\psi_n(\sigma),
\end{equation}
which obeys the initial condition $\tilde{\Pi}(0;\sigma,\sigma_0) = \delta(\sigma-\sigma_0)$ is satisfied.  Therefore, according to \eqref{eq:G_g_init} we have\footnote{Note that $P_{\text{\tiny FP},\mathrm{eq}}(\sigma_0) = \psi_0(\sigma_0)^2 = \psi_0(\sigma_0)\psi_0(\sigma)e^{4\pi^2V(\sigma)/3H^4}e^{-4\pi^2V(\sigma_0)/3H^4}$. 
}
\begin{align}\label{eq:temporal_corr_final}
    G_g(t) = \sum_n&\int\mathrm{d}\sigma_0 \psi_0(\sigma_0) g(\sigma_0) \psi_n(\sigma_0) e^{-\Lambda_nt}\nonumber \\
    \times &\int\mathrm{d}\sigma\psi_n(\sigma) g(\sigma) \psi_0(\sigma)= \sum_n g_n^2 e^{-\Lambda_nt},
\end{align}
where
\begin{equation}
\label{eq.g_n}
    g_n \equiv \int\mathrm{d}\sigma\psi_n(\sigma) g(\sigma) \psi_0(\sigma).
\end{equation}
We see that in late times the correlation is dominated by the smallest $\Lambda_n\neq 0$.

We can now present the equal-time correlation function by combining \eqref{eq:time_corr_to_spatial_corr} and \eqref{eq:temporal_corr_final}~\cite{Starobinsky:1986fx, Starobinsky:1994bd, Markkanen:2019kpv}:
\begin{equation}
    G_g(|\mathbf{x}_1 - \mathbf{x}_2|) = \sum_n \dfrac{g_n^2}{(aH|\mathbf{x}_1 - \mathbf{x}_2|)^{2\Lambda_n/H}}.
\end{equation}
We note that this depends on the physical distance between the two points at time $t$, namely, $a|\mathbf{x}_1 - \mathbf{x}_2|$.
This correlation function has the following dimensionless power spectrum~\cite{Markkanen:2019kpv},
\begin{align}
     &\Delta^2_g(k) = \dfrac{k^3}{2\pi^2} P_g(k) = \dfrac{k^3}{2\pi^2}\int\mathrm{d}^3r e^{-i\mathbf{k}\cdot\mathbf{r}}G_g(r) \nonumber\\
     \label{eq.g_n_expansion}
    & = \sum_n \dfrac{2 g_n^2}{\pi}\Gamma\left(2-\frac{2\Lambda_n}{H}\right)\sin\left(\frac{\pi\Lambda_n}{H}\right)\left(\dfrac{k}{aH}\right)^{2\Lambda_n/H} 
\end{align}
where $\Gamma$ denotes the gamma function. 
This expression is valid in the limit $k\ll aH$.
So far our discussion has been general and is valid for any potential under the slow-roll approximation and the assumption of a small effective mass, $|V''(\sigma)|\ll H^2$.
In the next section, we discuss a concrete example with $V(\sigma)$ given in \cref{eq:V_quartic}.

\section{Large Curvature Perturbation from Stochastic Fluctuations}
\label{sec:P_zeta}
We focus on the potential in \cref{eq:V_quartic} to demonstrate how large curvature perturbation can arise from stochastic fluctuations.
We first describe various equilibrium quantities and how to obtain the power spectra $P_{S_\sigma}$, and consequently evaluate $P_\zeta$ which determines the strength of the GW signal.
\subsection{Equilibrium Configuration}
The normalized PDF for the one-point function is given by \cref{eq:P_eq}. For convenience, we reproduce it here
\begin{align}
    P_{\rm FP,eq}(\sigma) = \frac{1}{\cal N} \exp\left(-\frac{8\pi^2 V(\sigma)}{3H^4}\right),
\end{align}
with
\begin{align}
    {\cal N} = \frac{2\sqrt{2}\sqrt{\lambda}}{\exp\left(\frac{m^4\pi^2}{3H^4\lambda}\right)m K_{{1\over 4}}\left(\frac{m^4\pi^2}{3H^4\lambda}\right)}.
\end{align}
Here $K_n(x)$ is the modified Bessel function of the second kind.
The mean displacement of the field can be computed as,
\begin{align}
    \langle \sigma^2\rangle = \int_{0}^{\infty}\D\sigma \sigma^2 P_{\rm FP,eq}(\sigma) = \frac{m^2}{2\lambda}\left(-1 + \frac{K_{{3\over 4}}\left(\frac{m^4\pi^2}{3H^4\lambda}\right)}{K_{{1 \over 4}}\left(\frac{m^4\pi^2}{3H^4\lambda}\right)} \right).
\end{align}
In the appropriate limits, this can be simplified to,
\begin{align}
    \langle \sigma^2\rangle\bigg\rvert_{\lambda\rightarrow 0} &=~ \frac{3H^4}{8\pi^2m^2},\\
    \langle \sigma^2\rangle\bigg\rvert_{m\rightarrow 0} &=~ \sqrt{\frac{3}{2\lambda}}\frac{\Gamma(3/4)}{\Gamma(1/4)\pi}H^2,
\end{align}
matching the standard results~\cite{Starobinsky:1994bd}.
We can also compute the average energy density of the field as,
\es{eq:V_sigma}{
\langle V(\sigma)\rangle &= \int_0^\infty \D\sigma V(\sigma) P_{\rm FP,eq}(\sigma) \\
&= \frac{1}{32}\left(\frac{3H^4}{\pi^2}-\frac{4m^4}{\lambda} + \frac{4m^4}{\lambda}\frac{K_{{3 \over 4}}\left(\frac{m^4\pi^2}{3H^4\lambda}\right)}{K_{{1\over 4}}\left(\frac{m^4\pi^2}{3H^4\lambda}\right)}\right),
}
reducing to,
\begin{align}
    \langle V(\sigma)\rangle \bigg\rvert_{\lambda\rightarrow 0} &=~ \frac{3H^4}{16\pi^2},\\
    \langle V(\sigma)\rangle \bigg\rvert_{m\rightarrow 0} &=~ \frac{3H^4}{32\pi^2}.
\end{align}
To ensure that $\sigma$ does not dominate energy density during inflation, we require
\begin{align}
	\langle V(\sigma) \rangle \ll 3 H^2 \mpl^2.
\end{align}
Finally, we compute $\langle V''(\sigma)\rangle$ to check the validity of slow-roll of the $\sigma$ field,
\es{}{
\langle V''(\sigma)\rangle &= \int_0^\infty \D\sigma V''(\sigma) P_{\rm FP,eq}(\sigma) \\
&= \frac{1}{2}m^2\left(-1 + \frac{3K_{{3\over 4}}\left(\frac{m^4\pi^2}{3H^4\lambda}\right)}{K_{{1\over 4}}\left(\frac{m^4\pi^2}{3H^4\lambda}\right)}\right),
}
which reduces to,
\begin{align}
    \langle V''(\sigma)\rangle \bigg\rvert_{\lambda\rightarrow 0} &=~ m^2,\\
    \langle V''(\sigma)\rangle \bigg\rvert_{m\rightarrow 0} &=~ \frac{3\sqrt{3}\Gamma(3/4)}{\sqrt{2}\pi \Gamma(1/4)}\sqrt{\lambda}H^2 \approx 0.4 \sqrt{\lambda}H^2.
\end{align}
To ensure slow-roll, we require
\begin{align}
\label{eq.slow-roll}
	\langle V''(\sigma)\rangle \ll H^2.
\end{align}

\subsection{Power Spectrum}
To obtain isocurvature power spectrum, $P_{S_\sigma}$, we need to compute the two-point function of $\delta\rho_\sigma/\rho_\sigma$.
We can write this more explicitly as,
\begin{align}
\label{eq.delta_rho}
	\frac{\delta\rho_\sigma(\vec{x})}{\rho_\sigma} = \frac{\rho_\sigma(\vec{x})-\langle\rho_\sigma(\vec{x})\rangle}{\langle\rho_\sigma(\vec{x})\rangle} = \frac{\rho_\sigma(\vec{x})}{\langle\rho_\sigma(\vec{x})\rangle} - 1.
\end{align}
where we can approximate $\rho_\sigma \approx V(\sigma)$, since $\langle V(\sigma) \rangle$ is approximately frozen, as long as~\cref{eq.slow-roll} is satisfied. 
Referring to~\cref{eq.g_n} and~\cref{eq.g_n_expansion}, the relevant coefficient $g_n$ for $\rho_\sigma$ is determined by, 
\begin{align}
\label{eq.g_n_quartic}
	g_n = \dfrac{\int \D\sigma \psi_n(\sigma) \rho_\sigma \psi_0(\sigma)}{\int \D\sigma \psi_0(\sigma) \rho_\sigma \psi_0(\sigma)}.
\end{align} 
For $n>0$, the last term in~\cref{eq.delta_rho} does not contribute because of the orthogonality of the eigenfunctions.

The eigenfunctions $\psi_n$ and the eigenvalues $\Lambda_n$ relevant for~\cref{eq.g_n_expansion} can be obtained by solving the eigensystem for the potential~\cref{eq:V_quartic}.
In terms of variables, $z = \lambda^{1/4}\sigma/H$ and $\alpha = m^2/(\sqrt{\lambda} H^2)$, the eigenvalue~\cref{eq:eigenvalue_eq} can be written as~\cite{Markkanen:2019kpv},
\es{}{
\frac{\partial^2\psi_n}{\partial z^2} &+ \left(-\left(\frac{4\pi^2}{3}\right)^2(\alpha z + z^3)^2 + \frac{4\pi^2}{3}(\alpha+3z^2)\right)\psi_n \\
&= - \frac{8\pi^2}{\sqrt{\lambda}}\frac{\Lambda_n}{H}\psi_n.
}
Given the potential in~\cref{eq:V_quartic}, the eigenfunctions are odd (even) functions of $\sigma$ for odd (even) values of $n$.
Since $\rho_\sigma$ is an even function of $\sigma$, \cref{eq.g_n_quartic} implies $g_1=0$, and therefore, the leading coefficient is $g_2$ with the eigenvalue $\Lambda_2$ determining the first non-zero contribution to the spectral tilt.
We show the numerical results for the eigenvalues for some benchmark parameter choices in Table~\ref{tab:eigen}.

\begin{center}
\begin{table}
	\begin{tabular}{| c | c | c | c | c | c |}
    \hline
    $m^2/H^2$ & $\lambda$ & $\Lambda_2/H$ & $g_2^2$ & $\Lambda_4/H$ & $g_4^2$ \\ 
    \hline
    \hline
    0.2 & 0.05 & 0.16 & $1.99$ & 0.37 & $0.03$ \\
    \hline
    0.2 & 0.07 & 0.17 & 1.98 & 0.40 & 0.05 \\
    \hline
    0.2 & 0.1 & 0.18 & $1.98$ & 0.44 & $0.07$ \\
    \hline
    0.25 & 0.05 & 0.19 & $1.99$ & 0.42 & $0.02$ \\
    \hline
    0.25 & 0.07 & 0.20 & 1.99 & 0.45 & 0.03 \\
    \hline
	0.25 & 0.1 & 0.21 & 1.98 & 0.49 & 0.05 \\
	\hline
	0.3 & 0.05 & 0.22 & $1.99$ & 0.48 & 0.01 \\
	\hline
	0.3 & 0.07 & 0.23 & 1.99 & 0.51 & 0.02 \\
	\hline
	0.3 & 0.1 & 0.24 & $1.99$ & 0.54 & $0.03$\\
	\hline
    \end{tabular}
    \caption{Eigenvalues for some benchmark parameter choices corresponding to the potential in~\cref{eq:V_quartic}.}
    \label{tab:eigen}
\end{table}
\end{center} 

The curvature power spectrum $\Delta^2_\zeta$ depends on both $\Delta^2_{S_\sigma}$ and $f_\sigma$, as in \cref{eq:Delta_zeta}.
With the values of $g_n, \Lambda_n$ in Table~\ref{tab:eigen}, we can compute the dimensionless power spectrum $\Delta^2_{S_\sigma}$ using \cref{eq.g_n_expansion}, where we can evaluate the factor of $aH$ at the end of inflation.
Furthermore, for our benchmark parameter choices, only the eigenvalue $\Lambda_2$ is relevant.
Therefore, \cref{eq.g_n_expansion} can be simplified as,
\es{eq:Delta_S}{
\Delta^2_{S_\sigma}(k) \approx \frac{2g_2^2}{\pi}\Gamma\left(2-{2\Lambda_2 \over H}\right) \sin\left({\pi \Lambda_2 \over H}\right)\left({k \over k_{\rm end}}\right)^{2\Lambda_2/H},
}
where $k_{\rm end} = a_{\rm end} H_{\rm end}$.

The precise value of $k_{\rm end}$ depends on the cosmological history after the CMB-observable modes exit the horizon.
It is usually parametrized as the number of $e$-foldings $N(k)\equiv \ln(a_{\rm end}/a_k)$, where $a_k$ is the scale factor when a $k$-mode exits the horizon during inflation, defined by $k = a_k H_k$.
Assuming an equation of state parameter $w$ between the end of inflation and the end of the reheating phase, we can derive the relation~\cite{Liddle:2003as, Dodelson:2003vq},
\es{}{
{k \over {a_0 H_0}} = \left({\sqrt{\pi}\over 90^{1/4}}{T_0 \over H_0}\right)e^{-N(k)} \left({V_k^{1/2} \over {\rho_{\rm end}^{1/4}\mpl}}\right) &\left({\rho_{\rm RH}\over \rho_{\rm end}}\right)^{{1-3w}\over {12(1+w)}}\\
&\times {g_{*,s,0}^{1/3}g_{*,\rm RH}^{1/4}\over g_{*,s,\rm RH}^{1/3}}.
}
Here $g_{*,\rm RH}$ and $g_{*,s,\rm RH}$ are the effective number of degrees of freedom in the energy density and entropy density, respectively, at the end of the reheating phase; $V_k$ is the inflationary energy density when the $k$-mode exits the horizon; $\rho_{\rm end}$ and $\rho_{\rm RH}$ are the energy densities at the end of inflation and reheating, respectively.
Plugging in the CMB temperature $T_0$ and the present-day Hubble parameter $H_0$, we arrive at
\es{eq:N_efolding}{
N(k) \approx 67 - \ln\left(\frac{k}{a_0H_0}\right) + \ln\left(\frac{V_k^{1/2}}{\rho_{\rm end}^{1/4} \mpl}\right) \\+ {{1-3w} \over {12(1+w)}} \ln\left(\frac{\rho_{\rm RH}}{\rho_{\rm end}}\right) + \ln\left(g_{*,\rm RH}^{1/4}\over g_{*,s,\rm RH}^{1/3}\right).
}
Significant sources of uncertainty in $N(k)$ comes from $V_k$, $\rho_{\rm end}$, $\rho_{\rm RH}$, and $w$.
Furthermore, \cref{eq:N_efolding} assumes a standard cosmological history where following reheating, the Universe becomes radiation dominated until the epoch of matter-radiation equality.
We now consider some benchmark choices with which we can evaluate $N(k)$.
We set $k=a_0H_0$, assume $V_k^{1/4} = 10^{16}$~GeV, close to the current upper bound~\cite{Planck:2018jri}, $\rho_{\rm end} \simeq V_k/100$, motivated by simple slow-roll inflation models, and $w\approx 0$~\cite{Abbott:1982hn, Dolgov:1982th, Albrecht:1982mp}.\footnote{The precise value of $w$ is model dependent, see, e.g.,~\cite{Podolsky:2005bw, Munoz:2014eqa, Lozanov:2016hid, Maity:2018qhi, Antusch:2020iyq} and~\cite{Allahverdi:2010xz} for a review.}
Then depending on the reheating temperature, we get
\es{eq:N_benchmark}{
N(k) =
\begin{cases}
	62, & T_{\rm RH} = 6\times 10^{15}~{\rm GeV},\\
	59, & T_{\rm RH} = 10^{11}~{\rm GeV}.
\end{cases}
}
For the first benchmark, we have assumed an instantaneous reheating after inflation, while for the second benchmark, the reheating process takes place for an extended period of time.
For these two benchmarks, $k_{\rm end} \approx 4\times 10^{23}~{\rm Mpc}^{-1}$ and $10^{22}~{\rm Mpc}^{-1}$, respectively.

To determine $\Delta_\zeta^2(k)$, we also need to evaluate $f_\sigma$ as a function of time. 
We can express the time dependence of $f_\sigma$ in terms of $k$ in the following way.
A given $k$-mode re-enters the horizon when $k = a_k H_k$, and assuming radiation domination, we get $k/k_{\rm end} =a_{\rm end}/a_k$. 
Since $f_\sigma$ increases with the scale factor before $\sigma$ decay, we can express $f_\sigma(t) = f_\sigma(t_d)(k_d/k)$, for $t<t_d$, where $k_d$ and $k$ are the modes that re-enter the horizon at time $t_d$ and $t$, respectively.
Therefore, the final expression for the curvature power spectrum at the time of mode re-entry follows from \cref{eq:Delta_zeta},
\es{eq:Delta_zeta_1}{
\Delta_\zeta^2(k) = 
\begin{cases}
	\Delta_{\zeta_r}^2(k) + \left(f_\sigma(t_d) \over {4+3 f_\sigma(t_d)}\right)^2 \Delta_{S_\sigma}^2(k),~k < k_d,\\
	 \Delta_{\zeta_r}^2(k) + \left(f_\sigma(t_d)(k_d/k) \over {4+3f_\sigma(t_d) (k_d/k)}\right)^2 \Delta_{S_\sigma}^2(k),~k > k_d.
\end{cases}
}

\begin{figure*}
    \centering
\includegraphics[width=0.33\textwidth]{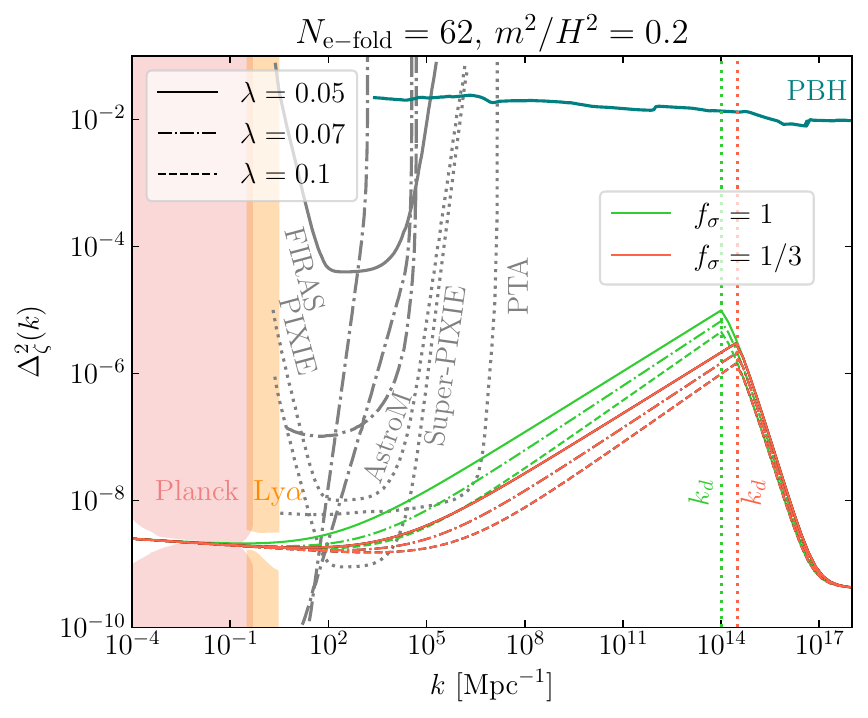}\hfill
    \includegraphics[width=0.33\textwidth]{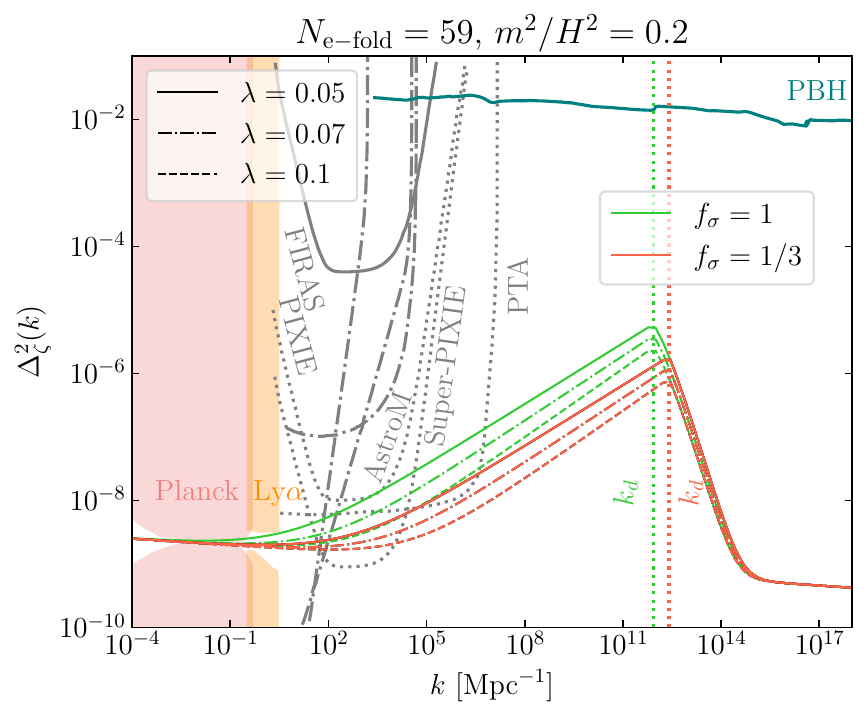}\hfill
    \includegraphics[width=0.33\textwidth]{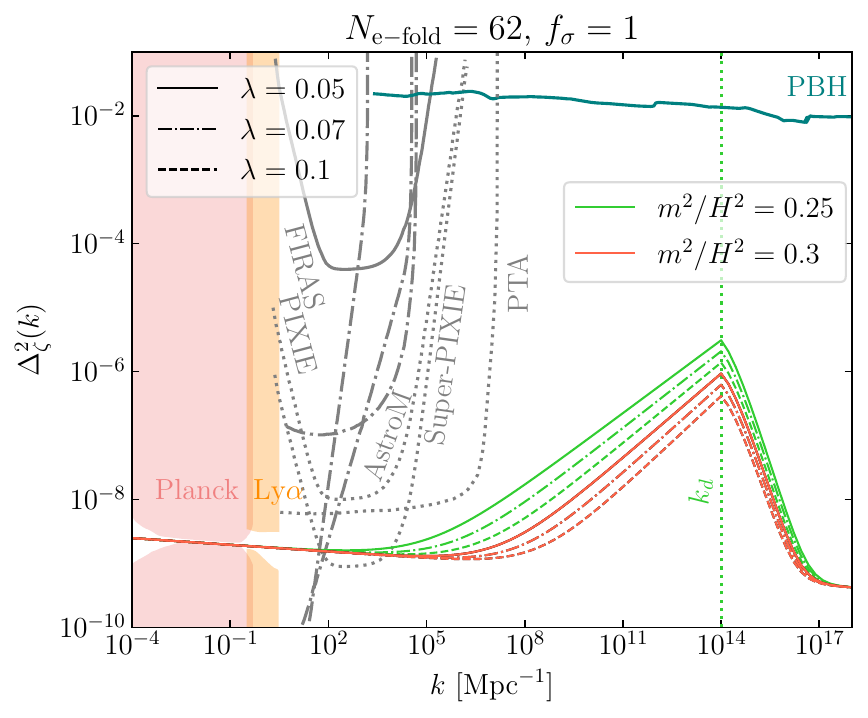}
    \caption{Power spectrum of curvature perturbations for the benchmarks discussed above. 
    Stochastic effects lead to a blue-tilted spectrum of $\sigma$, with larger $m$ and $\lambda$ corresponding to larger tilts, leading to faster decay as $k$ gets smaller. 
    The blue-tilt is eventually cut off at $k_d$, the $k$-mode that reenters the horizon at the time of $\sigma$ decay.
    For $k$ larger than $k_d$, the fractional energy density in $\sigma$ at the time of mode-reentry is smaller.
    Correspondingly, $\Delta_\zeta^2$ gets suppressed.
    Eventually, for very large $k$, the effects of $\sigma$ become negligible, and $\Delta_\zeta^2$ reverts back to its standard, slightly red-tilted behavior.
    A smaller value of $f_\sigma (k_d)$, the fractional energy density at the time $\sigma$ decay, suppresses the effect of $\sigma$ to $\Delta_\zeta^2$, and hence leads to a suppressed peak.
    This mechanism predicts signatures in CMB spectral distortion measurements~\cite{Chluba:2012gq}, especially in Super-PIXIE~\cite{Chluba:2019kpb}, along with Pulsar Timing Array (PTA) probes for enhanced DM substructure~\cite{Lee:2020wfn}, and precision astrometry probes (AstroM)~\cite{VanTilburg:2018ykj}. We also show constraints from FIRAS~\cite{2002ApJ...581..817F} and non-observation of primordial black holes (PBH)~\cite{Green:2020jor}.
    }
    \label{fig:power_spec}
\end{figure*}

To determine the scale $k_d$, we consider the benchmarks discussed above, along with some additional choices for other parameters.
\paragraph{Benchmark 1.}
We focus on the first benchmark in \cref{eq:N_benchmark}.
For $m^2 = 0.2 H^2$ and $\lambda \simeq 0.05 -0.1$, we get $\langle V(\sigma)\rangle \approx 0.02 H^4$ from \cref{eq:V_sigma}, implying $\langle V(\sigma)\rangle / V_k \approx 3 \times 10^{-12}$ for $H=5\times 10^{13}$~GeV.
Assuming instantaneous reheating, and $\rho_{\rm end}\simeq V_k/100$, we see $f_\sigma \simeq 1$ for $a \simeq (1/3)\times 10^{10} a_{\rm end}$.
As benchmarks, we assume $\sigma$ decays when $f_\sigma = 1$ and $1/3$.
Using $k_{\rm end}\approx 4\times 10^{23}~{\rm Mpc}^{-1}$, we can then evaluate  $k_d \approx 10^{14}~{\rm Mpc}^{-1}$ and $k_d \approx 3\times  10^{14}~{\rm Mpc}^{-1}$, respectively.
The result for the curvature power spectrum with these choices is shown in Fig.~\ref{fig:power_spec} (left).

\paragraph{Benchmark 2.}
We now discuss the second benchmark in \cref{eq:N_benchmark}.
We again choose $m^2 = 0.2 H^2$ and $\lambda \simeq 0.05 -0.1$, for which we get $\langle V(\sigma)\rangle \approx 0.02 H^4$ from \cref{eq:V_sigma}.
This implies $\langle V(\sigma)\rangle / V_k \approx 3 \times 10^{-12}$ for $H=5\times 10^{13}$~GeV, as before.
The rest of the parameters can be derived in an analogous way, with one difference.
During the reheating epoch, with our assumption $w\approx 0$, $f_\sigma$ does not grow with the scale factor since the dominant energy density of the Universe is also diluting as matter. 
Accounting for this gives $k_d \approx 8\times 10^{11}~{\rm Mpc}^{-1}$ and  $k_d \approx 3\times 10^{12}~{\rm Mpc}^{-1}$, for $f_\sigma =1 $ and $1/3$, respectively, with the resulting curvature power spectrum shown in Fig.~\ref{fig:power_spec} (center).
\paragraph{Benchmark 3.} This is same as the first benchmark discussed above, except we focus on $m^2 = 0.25 H^2$ and $0.3H^2$ along with $f_\sigma = 1$.
The result is shown in Fig.~\ref{fig:power_spec} (right).

\section{Gravitational Wave Signature}
\label{sec:GW}
\subsection{Secondary Gravitational Waves from Scalar Curvature Perturbation}
We now review how large primordial curvature perturbations can source GW at the second order in perturbation theory \cite{Ananda:2006af,Baumann:2007zm} (for a review see~\cite{Domenech:2021ztg}).
We then evaluate the GW spectrum sourced by $\Delta_\zeta^2$ computed in~\cref{sec:P_zeta}.
We start our discussion with a brief review of the essential relations and expand the discussion further in \cref{app:GW_details}.

We can write a tensor perturbation in Fourier space as,
\begin{equation}
    h_{ij}(\tau,\mathbf{x})=\sum_{\lambda=+,\times}\int\dfrac{\mathrm{d}^3k}{(2\pi)^{3}}e^{i\mathbf{k}\cdot\mathbf{x}}\epsilon_{ij}^\lambda(\mathbf{k})h_\lambda(\tau,\mathbf{k})\,,
\end{equation}
where $\epsilon_{ij}^{\lambda=\{+,\times\}}(\mathbf{k})$ are polarization tensors:
\begin{align}
    \epsilon_{ij}^{+}(\mathbf{k}) = &\dfrac{1}{\sqrt{2}}\left(\mathrm{e}_{1,i}(\mathbf{k})\mathrm{e}_{1,j}(\mathbf{k})-\mathrm{e}_{2,i}(\mathbf{k})\mathrm{e}_{2,j}(\mathbf{k})\right),\\
    \epsilon_{ij}^{\times}(\mathbf{k}) = &\dfrac{1}{\sqrt{2}}\left(\mathrm{e}_{1,i}(\mathbf{k})\mathrm{e}_{2,j}(\mathbf{k})+\mathrm{e}_{2,i}(\mathbf{k})\mathrm{e}_{1,j}(\mathbf{k})\right),
\end{align}
with $\mathrm{e}_{1,2}$ the orthonormal bases spanning the plane transverse to $\mathbf{k}$. 
The equation of motion determining the generation and evolution of GW is given by
\begin{equation}\label{eq:h_eom}
    h_\lambda''(\tau,\mathbf{k})+2\mathcal{H} h_\lambda'(\tau,\mathbf{k})+k^2h_\lambda(\tau,\mathbf{k})=4\mathcal{S}_\lambda(\tau,\mathbf{k}),
\end{equation}
where $'$ denotes derivative with respect to the conformal time $\tau$ and $\mathcal{H}=a'/a$ is the conformal Hubble parameter. 
The second-order (in scalar metric perturbation $\Phi$) source term is given by\footnote{We parametrize the scalar metric fluctuations, for vanishing anisotropic stress, as \begin{equation}
    \mathrm{d} s^2 = -\left(1+2\Phi\right)\mathrm{d}t^2 + a^2\left(1-2\Phi\right)\delta_{ij} \mathrm{d} x^i\mathrm{d} x^j
\end{equation}} 
\es{eq:source_Bardeen}{
&\mathcal{S}_\lambda(\tau,\mathbf{k})=
\int\dfrac{\mathrm{d}^3q}{(2\pi)^{3}} \dfrac{Q_\lambda(\mathbf{k},\mathbf{q})}{3(1+w)}\bigg[2(5+3w)\Phi_{\mathbf{p}}\,\Phi_{\mathbf{q}}\\
&+\tau^2(1+3w)^2\Phi_{\mathbf{p}}'\,\Phi_{\mathbf{q}}'+2\tau(1+3w)(\Phi_{\mathbf{p}}\,\Phi_{\mathbf{q}}'+\Phi_{\mathbf{p}}\,\Phi_{\mathbf{q}}')\bigg].   
}
We have defined $\mathbf{p}\equiv\mathbf{k}-\mathbf{q}$, $\Phi_{\mathbf{k}}\equiv\Phi(\tau,\mathbf{k})$, and a projection operator $Q_\lambda(\mathbf{k},\mathbf{q})$: 
\begin{equation}\label{eq:Q_definition}
    Q_\lambda(\mathbf{k},\mathbf{q})\equiv \epsilon_\lambda^{ij}(\mathbf{k})q_iq_j\,.
\end{equation}
The metric perturbation $\Phi(\tau,\mathbf{k})$ can be written in terms of the primordial curvature perturbation $\zeta(\mathbf{k})$,
\begin{equation}\label{eq:bardeen_intermsof_transferfunc}
    \Phi(\tau,\mathbf{k})=\dfrac{3+3w}{5+3w}T_\Phi(k\tau)\zeta(\mathbf{k})\,,
\end{equation}
via a transfer function $T_\Phi(k\tau)$ which depends on $w$.
With the above quantities, one can now solve \cref{eq:h_eom} using the Green function method,\footnote{Scale factors appearing in the $I$ integral as $a(\bar{\tau})/a(\tau)$ are the artifact of $G_{\mathbf{k}}(\tau,\bar{\tau})$ being Green's function of the new variable $v(\tau,\mathbf{k})=a h(\tau,\mathbf{k})$ and not $h_\lambda$ itself; see Appendix \ref{app:greensfunc}.} 
\begin{equation}\label{eq:GW_solution}
    h_\lambda(\tau,\mathbf{k})=\dfrac{4}{a(\tau)}\int_{\tau_0}^{\tau}\mathrm{d}\bar{\tau} G_{\mathbf{k}}(\tau,\bar{\tau}) a(\bar{\tau})\mathcal{S}_\lambda(\bar{\tau},\mathbf{k})\,.
\end{equation}
Using the solutions of \cref{eq:h_eom}, the power spectrum $ P_{\lambda}(\tau,k)$, defined via,  
\begin{equation}
    \langle h_{\lambda_1}(\tau,\mathbf{k}_1)h_{\lambda_2}(\tau,\mathbf{k}_2)\rangle \equiv (2\pi)^{3} \delta_{\lambda_1\lambda_2}\delta^3(\mathbf{k}_1+\mathbf{k}_2) P_{\lambda_1}(\tau,k_1)\,,
\end{equation}
can be written as,
\es{eq:master}{
    &\langle h_{\lambda_1}(\tau,\mathbf{k}_1)h_{\lambda_2}(\tau,\mathbf{k}_2)\rangle = \\
    &16\int\dfrac{\mathrm{d}^3q_1}{(2\pi)^{3}}\dfrac{\mathrm{d}^3q_2}{(2\pi)^{3}} Q_{\lambda_1}(\mathbf{k}_1,\mathbf{q}_1)Q_{\lambda_2}(\mathbf{k}_2,\mathbf{q}_2)I(|\mathbf{k}_1-\mathbf{q}_1|,q_1,\tau_1) \\
    &\times\,I(|\mathbf{k}_2-\mathbf{q}_2|,q_2,\tau_2)\langle\zeta(\mathbf{q}_1)\zeta(\mathbf{k}_1-\mathbf{q}_1)\zeta(\mathbf{q}_2)\zeta(\mathbf{k}_2-\mathbf{q}_2)\rangle\,.
}
Here
\begin{equation}\label{eq:I_time_integral}
    I(p,q,\tau)=\dfrac{1}{a(\tau)}\int_{\tau_0}^\tau\mathrm{d}\bar{\tau}~ G_{\mathbf{k}}(\tau,\bar{\tau})a(\bar{\tau})f(p,q,\bar{\tau})\,,
\end{equation}
and
\es{eq:f_function}{
    &\dfrac{(5+3w)^2}{3(1+w)}f(p,q,\tau)= 2(5+3w)T_\Phi(p\tau)\,T_\Phi(q\tau)\\
    &+\tau^2(1+3w)^2T'_\Phi(p\tau)\,T'_\Phi(q\tau)\\
    &+2\tau(1+3w)\left[T_\Phi(p\tau)\,T'_\Phi(q\tau)+T'_\Phi(p\tau)\,T_\Phi(q\tau)\right].
}
where $T'_\Phi(p\tau) = \partial T_\Phi(p\tau)/\partial\tau$.
We note that the power spectrum is sourced by the four-point correlation function of super-horizon curvature perturbations, and is further modified by the sub-horizon evolution as encapsulated in $I(p,q,\tau)$.   

The four-point function in \cref{eq:master} has both disconnected and connected contributions, from the scalar power spectrum and trispectrum, respectively. 
The connected contribution usually contributes in a subdominant way compared to the disconnected piece in determining total GW energy density; see \cite{Garcia-Saenz:2022tzu} for a general argument.\footnote{ See also \cite{Adshead:2021hnm, Unal:2018yaa, Atal:2021jyo} for examples where the connected contribution can be important.} 
Therefore, in the following, we focus only on the disconnected contribution, which can be written as
\es{eq:GW_power_disconnected}{
     P_\lambda(\tau,k)\bigg|_\mathrm{d}  &= 32\int\dfrac{\mathrm{d}^3q}{(2\pi)^{3}} Q_{\lambda}(\mathbf{k},\mathbf{q})^2I(|\mathbf{k}-\mathbf{q}|,q,\tau)^2  \\
     &\times P_\zeta(q) P_\zeta(|\mathbf{k}-\mathbf{q}|)\,.
}
For a derivation of this formula see \cref{app:conn_disconn}.

GW signal strength can be characterized by SGWB energy density per unit logarithmic interval of frequency and normalized to the total energy density \cite{Maggiore:1999vm},
\es{eq:om_gw}{
h^2\Omega_\text{\tiny{GW}} = \dfrac{1}{\rho_{\rm tot}}\dfrac{{\rm d}\rho_\text{\tiny{GW}}}{{\rm d}\log f}
}
where the present day Hubble parameter is given by $H_0 =100h{\rm~km/s/Mpc} $ and $\rho_{\rm tot} = 3 \mpl^2 H_0^2$ is the critical energy density in terms of the reduced Planck mass $\mpl \approx 2.4\times 10^{18}$~GeV. 
The total energy density $\rho_\text{\tiny{GW}}$ is given by,
\es{eq:rho_GW_tot}{
&\rho_\text{\tiny{GW}} = {\mpl^2 \over 4} \int \D\ln k {k^3 \over {16\pi^2}} \times 
\\
&\sum_\lambda\left( \langle {\dot{h}_\lambda(t,{\bf{k}}) \dot{h}_\lambda(t,-{\bf{k}})}\rangle' + {k^2\over a^2} \langle {h_\lambda(t,{\bf{k}}) h_\lambda(t,-{\bf{k}})}\rangle' \right),
}
with the primes denoting the fact that momentum-conserving delta functions are factored out, $\langle {h_\lambda(t,{\bf{k}}) h_\lambda(t,{\bf{k}'})}\rangle= (2\pi)^3 \delta^3({\bf k}+{\bf k}') \langle {h_\lambda(t,{\bf{k}}) h_\lambda(t,-{\bf{k}})}\rangle'$.
Approximating $\dot{h}_\lambda(t,{\bf{k}}) \approx (k/a) h_\lambda(t,{\bf{k}})$, we can simplify to get,\footnote{Note that we are using the convention at which the spatial part of the metric is given by $a^2(\delta_{ij}+h_{ij}/2)\mathrm{d}x^i\mathrm{d}x^j$. If we were using an alternative convention $a^2(\delta_{ij}+h_{ij})\mathrm{d}x^i\mathrm{d}x^j$, then the factor of $1/48$ would be replaced by $1/12$ as in refs.~\cite{Maggiore:1999vm,Kohri:2018awv}.}
\es{eq:GW_energy_density}{
\Omega_\text{\tiny{GW}} = \dfrac{1}{48}\left(\dfrac{k}{a(\tau)H(\tau)}\right)^2 \sum_{\lambda=+,\times}\Delta_\lambda^2(\tau,k),
}
where $\Delta_\lambda^2(\tau,k) = (k^3/(2\pi^2)) P_\lambda(\tau,k)$.

\begin{figure}[htbp]
    \centering
    \begin{overpic}[width=0.49\textwidth]{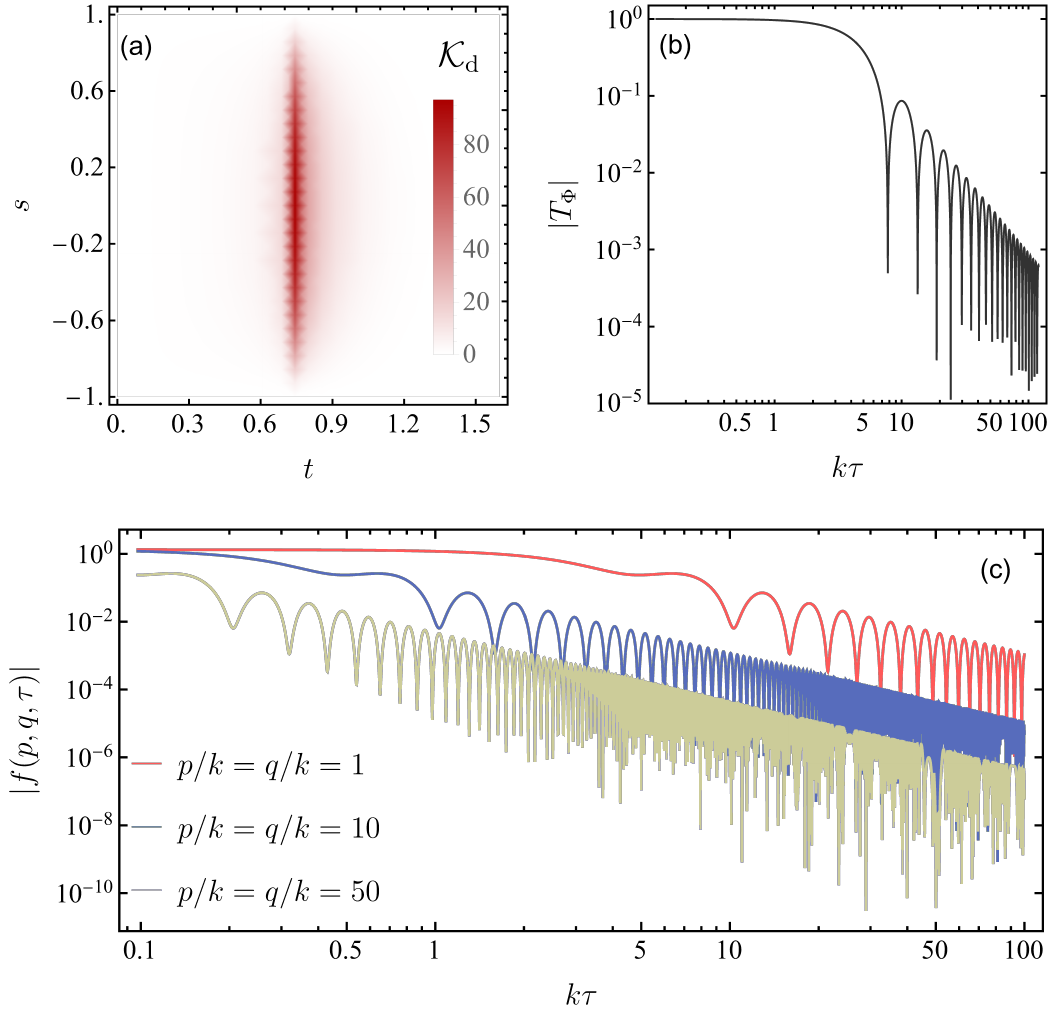}
\end{overpic} 
    \caption{(a) The kernel function from eq.~\eqref{eq:kernel}. We note a clear resonance contribution from $t\simeq0.7$ corresponding to $u+v\simeq\sqrt{3}$. (b) The transfer function $T_\Phi$. (c) Function $f(p,q,\tau)$ as in eq. \eqref{eq:f_function}. We see that for the scalar modes that enter the horizon earlier, with $p,q>k$, this function is more suppressed as expected from the behavior of the transfer function. }
    \label{fig:kernel}
\end{figure}

The above expression can be rewritten in form convenient for numerical evaluation
(see appendix \ref{app:recast_numeric} for a derivation),\footnote{Note that the integration variable $u$ and $v$ are swapped with $t$ and $s$ since in the $t-s$ space, integration limits are independent of the integration variables.} 
\begin{align}
    \Omega_{\text{\tiny{GW}}}(k) = \dfrac{2}{48\alpha^2}\int_0^\infty\mathrm{d} t\int_{-1}^{1}\mathrm{d} s ~\mathcal{K}_{\mathrm{d}}(u,v)  \Delta^2_\zeta(uk) \Delta^2_\zeta(vk) \label{eq:kernel}
\end{align}
where $u=|\mathbf{k}-\mathbf{q}|/k=p/k, v=q/k, s=u-v, t=u+v-1$, and $\mathcal{K}_\mathrm{d}$ is the kernel function following from manipulating the integrand of eq.~\eqref{eq:GW_power_disconnected}.
This kernel function is illustrated in \cref{fig:kernel}a.

We now focus on the scenario where GW is generated during a radiation dominated epoch and set $w=1/3$.
We can then write (see Appendix~\ref{app:transfer_func} for details),
\begin{equation}\label{eq:transfer_func_RD}
    T_\Phi(k\tau)=\frac{9 \sqrt{3}}{(k \tau)^3}\left(\sin \frac{k \tau}{\sqrt{3}}-\frac{k \tau}{\sqrt{3}} \cos \frac{k \tau}{\sqrt{3}}\right)\,,
\end{equation}
and plot this function in \cref{fig:kernel}b.
We note that after entering the horizon, modes start to oscillate and decay, and as a result, the sub-horizon modes do not significantly contribute to GW generation. 
In \cref{fig:kernel}c, we confirm that at any given time $f(p,q,\tau)$ is suppressed for shorter modes that have re-entered the horizon earlier.
Finally, the green function is given by (see \cref{app:greensfunc} for details)
\begin{equation}
\label{eq:green}
    G_{\mathbf{k}}(\tau,\bar{\tau})=\dfrac{\sin [k(\tau-\bar{\tau})]}{k}\,.
\end{equation}

With these expressions, we can obtain a physical understanding of GW generation via \cref{eq:GW_power_disconnected}.
The Green function, given in \cref{eq:green}, is an oscillatory function of time whose frequency is $k$. 
The quantity $f(p,q,\tau)$ is also an oscillatory and decaying function of time (see \cref{fig:kernel}c), inheriting these properties from the transfer function \eqref{eq:transfer_func_RD}. 
Therefore, the dominant contribution to the integral \eqref{eq:I_time_integral} is a resonant contribution when the momentum of the produced GW is of the same order as the momentum of the scalar modes, i.e., $k\sim p\sim q$. In particular, the resonant point is at $u+v\simeq\sqrt{3}$ \cite{Garcia-Saenz:2022tzu} as shown in \cref{fig:kernel}a. 
GW generation is suppressed in other parts of the phase space. 
For example, the source term, which contains gradients of the curvature perturbation \cite{Baumann:2007zm}, is suppressed by small derivatives if any of the wavenumbers $p, q$ of $\zeta$ is much smaller than $k$.
On the other hand, if $p, q$ are much larger than $k$, then the scalar modes would have decayed significantly after entering the horizon by the time $k\sim H$, and thus the production of GW with momentum $k$ gets suppressed. 

To obtain the final result for $\Omega_{\rm GW}$, we note that the GW comoving wavenumber $k$ is related to the present-day, redshifted frequency $f$ of the generated GW via
\begin{equation}
    f = f_\ast \left(\dfrac{a_\ast}{a_0}\right) = \dfrac{k}{2\pi} \simeq 1.5\,\mathrm{mHz} \left(\dfrac{k}{10^{12}\,\mathrm{Mpc^{-1}}}\right),
\end{equation}
where $f_\ast$ and $a_\ast$ are respectively the frequency and the scale factor at the time of GW generation.
\begin{figure*}
    \centering
    \includegraphics[width=0.33\textwidth]{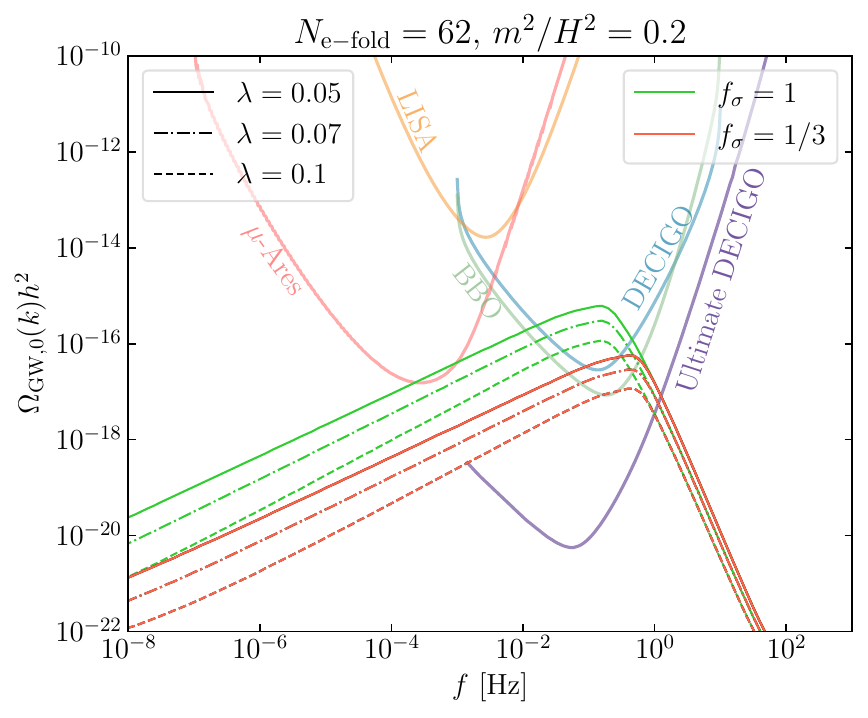}\hfill
    \includegraphics[width=0.33\textwidth]{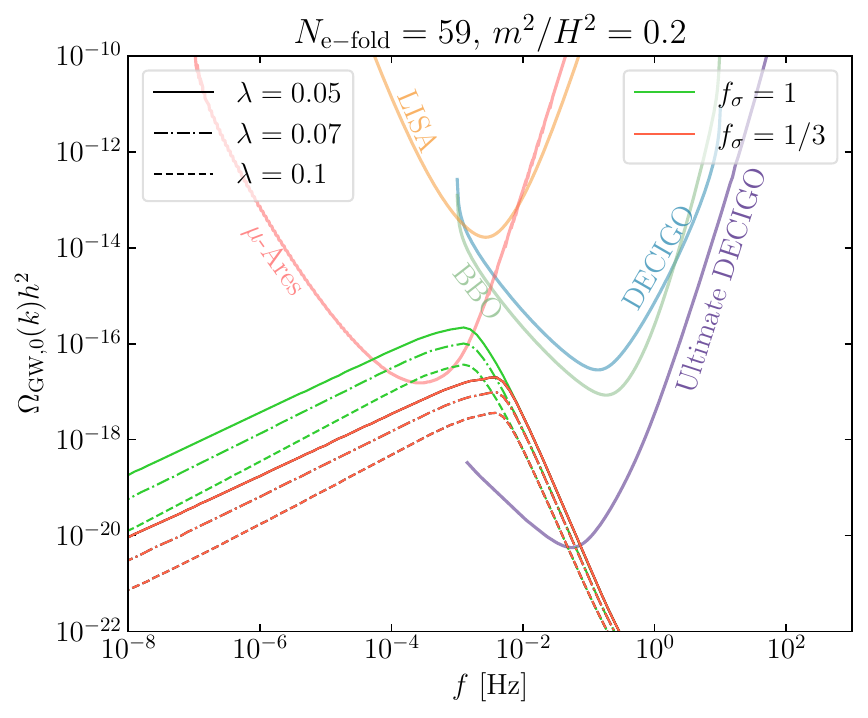}\hfill
    \includegraphics[width=0.33\textwidth]{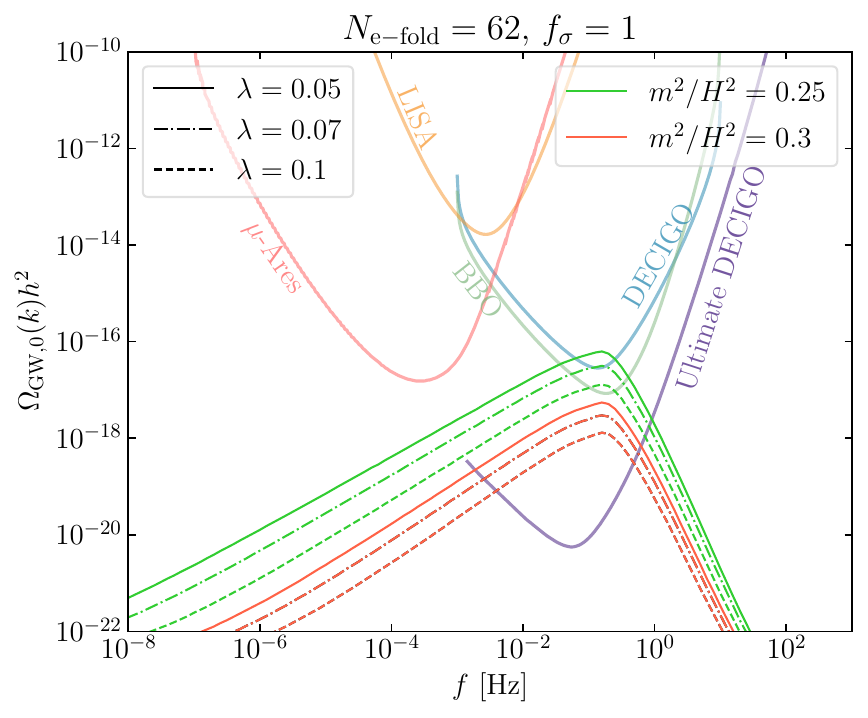}
    \caption{Gravitational wave spectrum for the benchmarks discussed in Fig.~\ref{fig:power_spec}. We notice that the number of $e$-folds after CMB-observable modes exited the horizon determines the peak frequency of the spectrum, and correspondingly, different detectors can be sensitive to the signal. Although a similarly peaked spectrum would appear in the context of cosmological phase transitions (PT), the low-frequency tail of this GW spectrum is different from the usual $f^3$ tail. While in the context of PT the $f^3$ scaling originates due to causality and superhorizon behavior of fluctuations, in our scenario, the $f$-scaling is determined by $\sigma$ mass. The differing frequency dependence can then be used to discriminate between the two classes of signals. 
    }
    \label{fig:gw}
\end{figure*}
Using these expressions, we arrive at our final result, shown in Fig.~\ref{fig:gw}, for the same benchmark choices discussed in Fig.~\ref{fig:power_spec}.
We see that stochastic effects can naturally give rise to a large enough SGWB, within the sensitivity range of DECIGO, BBO, $\mu$-Ares, and Ultimate DECIGO \cite{Schmitz:2020syl, Sesana:2019vho, Braglia:2021fxn}.

\section{Conclusion}
\label{sec:concl}
In this work, we have discussed an early Universe scenario containing a light spectator field, along with an inflaton field.
The fluctuations of the inflaton are red-tilted and explain the observed fluctuations in the CMB and LSS.
On the other hand, the spectator field $\sigma$ naturally acquires a blue-tilted power spectrum.
This blue-tilted power spectrum is eventually cut-off at very small scales since when such small-scale modes enter the horizon, the spectator field contributes subdominantly to the total energy density.
As a consequence, primordial black holes are not produced in this scenario.
Overall, this mechanism of generating a blue-tilted spectrum works for any generic inflaton potential and does not require any particular fine-tuning or structure such as an inflection point or a bump on the potential or an ultra slow-roll phase.

The blue-tilted spectrum gives rise to large curvature perturbations at small scales.
These, in turn, source a stochastic gravitational wave background (SGWB) when the perturbations re-enter the horizon.
Focusing on some benchmark choices for the number of $e$-foldings and spectator field potential, we have shown that this scenario predicts observable gravitational waves at future detectors operating in $10^{-5}$~Hz to $10$~Hz range, with strengths $\Omega_{\rm GW}h^2 \simeq 10^{-20}-10^{-15}$.

There are various interesting future directions.
In particular, we have worked in a regime where $\sigma$ does not dominate the energy density during the cosmological history.
It would be interesting to explore the consequences of an early matter-dominated era caused by the $\sigma$ field.
We have also seen that the low-frequency scaling of the SGWB spectrum depends on the mass and coupling of $\sigma$ and is generally different from the $f^3$-scaling expected in the context of cosmological PT, or $f^{2/3}$-scaling expected in the context of binary mergers.
This different frequency dependence can be used to identify the origin of an SGWB, and distinguish between various cosmological or astrophysical contributions.
Along these lines, it would be interesting to carry out a quantitative analysis to understand how well we can separate any two frequency dependencies, for example, by doing a Fisher analysis.
\section*{Note Added}
While we were finishing this work, the NANOGrav result combining 15-year data appeared~\cite{NANOGrav:2023gor}.  Secondary gravitational waves from the scalar perturbation can in principle give rise to the signal~\cite{NANOGrav:2023hvm}.
Such scalar perturbations could be generated in a model similar to the one considered in this paper. 
However, the frequency dependence of $\Omega_{\rm GW}h^2$ determined by the NANOGrav result is~\cite{NANOGrav:2023gor} $1.8\pm 0.6$.
We note that for a free field with mass $m$, the frequency dependence of $\Omega_{\rm GW}h^2$ is given by, $4m^2/(3H^2)$. So for the central value, one would naively infer $m^2/H^2 = 1.4$. 
Therefore to interpret it in terms of a free field, we require a mass bigger than the Hubble scale. However, since for larger than Hubble-scale masses, the stochastic effects are not efficient, one may have to go beyond the stochastic scenario to explain the NANOGrav observations.
We could instead consider a regime in which the misalignment contribution is important~\cite{Kasuya:2009up, Kawasaki:2012wr}. We will leave a detailed analysis of this scenario to future work.

\section*{Acknowledgment}
We thank Keisuke Harigaya, Andrew Long, and Neal Weiner for their helpful discussions.
RE is supported in part by the University of Maryland Quantum Technology Center. SK is supported in part by the National Science Foundation (NSF) grant PHY-1915314 and the U.S. DOE Contract DE-AC02-05CH11231.
SK thanks Aspen Center for Physics, supported by NSF grant PHY-2210452, for hospitality while this work was in progress. 
The research of AM is supported by the U.S. Department of Energy, Office of Science,
Office of Workforce Development for Teachers and Scientists, Office of Science Graduate Student Research (SCGSR) program under contract number DE‐SC0014664. LTW is supported by the DOE
grant DE-SC0013642. 

\appendix
\section{Scalar-induced gravitational waves: technical details}\label{app:GW_details}
\subsection{Transfer functions}\label{app:transfer_func}
The equation of motion for the scalar perturbation $\Phi$ in the absence of isocurvature perturbations is,
\begin{equation}
    \Phi''(\tau,\mathbf{k})+3(1+c_s^2)\mathcal{H}\Phi'(\tau,\mathbf{k})+c_s^2 k^2\Phi(\tau,\mathbf{k})=0\,,
\end{equation}
where $c_s^2\simeq w$ is the sound speed of the fluid. Defining dimensionless parameter $y=\sqrt{w}k\tau$, we rewrite this equation as
\begin{equation}
    \dfrac{\mathrm{d}^2\Phi(y,\mathbf{k})}{\mathrm{d} y^2}+\dfrac{6(1+w)}{1+3w}\dfrac{1}{y}\dfrac{\mathrm{d}\Phi(y,\mathbf{k})}{\mathrm{d} y}+\Phi(y,\mathbf{k}) = 0\,.
\end{equation}
A general solution is given by,
\begin{equation}\label{eq:bardeen_solution}
    \Phi(y,\mathbf{k}) = y^{-\gamma}\left[C_1(\mathbf{k})J_\gamma(y)+C_2(\mathbf{k})Y_\gamma(y)\right]\,,
\end{equation}
where $J_\gamma$ and $Y_\gamma$ are spherical Bessel functions of the first and second kind, respectively, of order $\gamma$
\begin{equation}
    \gamma=\dfrac{3(1+w)}{1+3w}-1\,.
\end{equation}

In the radiation dominated era, in which $w=1/3\rightarrow\gamma=1$, we have
\begin{align}
    \Phi(y,\mathbf{k}) = \dfrac{1}{y^2}\bigg[&C_1(\mathbf{k})\left(\dfrac{\sin y}{y}-\cos y\right)+\nonumber\\
    &C_2(\mathbf{k})\left(\dfrac{\cos y}{y}+\sin y\right)\bigg]\,.
\end{align}
We can deduce the initial conditions of this solution by considering the early-time limit $k\tau\ll1$,
\begin{equation}
    \dfrac{\sin y}{y}-\cos y \simeq \dfrac{y^2}{3}\quad\text{and}\quad\dfrac{\cos y}{y}+\sin y \simeq \dfrac{1}{y}\,.
\end{equation}
The first term ($\propto C_1$) is then constant in this limit, while the second term ($\propto C_2$) decays as $1/y^3 \sim 1/a^3$. We can therefore assume the initial conditions, 
\begin{equation}\label{eq:bardeen_initial_conditions}
    C_1(\mathbf{k})= 2 \zeta(\mathbf{k}),\quad C_2(\mathbf{k})=0\,,
\end{equation}
which gives a particular solution,
\begin{equation}
    \Phi(\tau,\mathbf{k}) = \frac{2}{3} \zeta(\mathbf{k})\dfrac{3}{y^2}\left(\dfrac{\sin y}{y}-\cos y\right)\,,
\end{equation}
resulting in the transfer function, via \eqref{eq:bardeen_intermsof_transferfunc},
\begin{equation}
    T_\Phi(k\tau)=\frac{3}{(k \tau / \sqrt{3})^3}\left(\sin \frac{k \tau}{\sqrt{3}}-\frac{k \tau}{\sqrt{3}} \cos \frac{k \tau}{\sqrt{3}}\right)\,.
\end{equation}
We can now see the distinct behavior of super-horizon ($k\tau\ll1$) and sub-horizon ($k\tau\gg1$) modes in the radiation dominated era. While the super-horizon modes freeze via our analysis above, the sub-horizon modes oscillate and damp as $\sim\cos k\tau/(k\tau)^2$.

In the matter dominated era, $w=0$ and the equation of motion for $\Phi$ becomes, 
\begin{equation}
\Phi''(\tau,\mathbf{k})+3\mathcal{H}\Phi'(\tau,\mathbf{k})=0\,,
\end{equation}
leading to a constant transfer function.

\subsection{Green's function and GW solution}\label{app:greensfunc}
In this subsection, we discuss in detail the solutions to \cref{eq:h_eom}, which is derived using the second-order Einstein equation, $G^{(2)}_{ij}=8\pi G T^{(2)}_{ij}$, for second-order tensor and first-order scalar contributions. We neglect scalar anisotropic stress, and second-order vector and scalar perturbations. In other words, we use the following perturbed FLRW metric in the Newtonian gauge,
\begin{equation}
    \mathrm{d} s^2 = -\left(1+2\Phi\right)\mathrm{d}t^2 + a^2\left(\left(1-2\Phi\right)\delta_{ij}+\dfrac{1}{2}h_{ij}\right) \mathrm{d} x^i\mathrm{d} x^j,
\end{equation}
assuming a perfect fluid energy-momentum tensor with equation of state $w$. Using lower order solutions and projecting out spatial indices using polarization tensors, i.e. $\epsilon_\lambda^{ij}T_{ij}=T_\lambda$ for any tensor $T$, we recover \eqref{eq:h_eom}.
For simplicity, we define a new variable $v(\tau,\mathbf{k})=ah_\lambda(\tau,\mathbf{k})$, which gives the equation of motion for $v(\tau,\mathbf{k})$,
\begin{equation}\label{eq:eom_of_v}
    v''(\tau,\mathbf{k})+\left[k^2-\dfrac{a''(\tau)}{a(\tau)}\right]v(\tau,\mathbf{k})=4a(\tau)\mathcal{S}_\lambda(\tau,\mathbf{k})\,.
\end{equation}
We need the two homogeneous solutions of this equation $v_1(\tau)$ and $v_2(\tau)$ to construct the Green's function,
\begin{equation}\label{eq:Greenfunc_generic}
    G_{\mathbf{k}}(\tau,\bar{\tau})=\dfrac{v_1(\tau)v_2(\bar{\tau})-v_1(\bar{\tau})v_2(\tau)}{v_1'(\bar{\tau})v_2(\bar{\tau})-v_1(\bar{\tau})v_2'(\bar{\tau})}\,.
\end{equation}
For each $\mathbf{k}$ we have
\begin{align}
    v_{1,2}''(\tau)+\left[k^2-\dfrac{a''(\tau)}{a(\tau)}\right]v_{1,2}(\tau)=0
\end{align}
which, using $a\propto \tau^\alpha$ and $x=k\tau$, leads to
\begin{align}\label{eq:homogeneous_eomofv}
    \dfrac{\mathrm{d}^2v_{1,2}(x)}{\mathrm{d} x^2}+\left[1-\dfrac{\alpha(\alpha-1)}{x^2}\right]v_{1,2}(x)=0\,,
\end{align}
where $\alpha=2/(1+3w)$. The solutions are
\begin{align}
    v_1(x)=\sqrt{x}J_{\alpha-1/2}(x)\,\\
    v_2(x)=\sqrt{x}Y_{\alpha-1/2}(x)\,
\end{align}
where $J_{\alpha-1/2}$ and $Y_{\alpha-1/2}$ are again spherical Bessel functions of first and second kind, respectively. We note that
\begin{align}
    \dfrac{\mathrm{d} v_1}{\mathrm{d} x}=\dfrac{\alpha}{\sqrt{x}}J_{\alpha-1/2}(x)-\sqrt{x}J_{\alpha+1/2}\,\\
    \dfrac{\mathrm{d} v_2}{\mathrm{d} x}=\dfrac{\alpha}{\sqrt{x}}Y_{\alpha-1/2}(x)-\sqrt{x}Y_{\alpha+1/2}\,.
\end{align}
Now, we can calculate the expression in the denominator of the Green's function,
\begin{align}
    v_1'(x)v_2(x)-v_1(x)v_2'(x)&=kx\bigg[J_{\alpha-1/2}(x)Y_{\alpha+1/2}(x)-\nonumber\\
    &~~~~~~~~~J_{\alpha+1/2}(x)Y_{\alpha-1/2}(x)\bigg]\nonumber\\
    &=-\dfrac{2}{\pi}\,.
\end{align}
The second equality can be checked explicitly via \texttt{Mathematica}. Thus, \eqref{eq:Greenfunc_generic} simplifies to
\begin{align}
    G_{\mathbf{k}}(\tau,\bar{\tau})=\dfrac{\pi}{2}\sqrt{\tau\bar{\tau}}\bigg[&J_{\alpha-1/2}(k\bar{\tau})Y_{\alpha-1/2}(k\tau)-\nonumber\\
    &J_{\alpha-1/2}(k\tau)Y_{\alpha-1/2}(k\bar{\tau})\bigg]\,.
\end{align}

In the radiation dominated era, $\alpha=1$, and so,
\begin{equation}
    G_{\mathbf{k}}(\tau,\bar{\tau})=\dfrac{\sin k(\tau-\bar{\tau})}{k}\,,
\end{equation}
where we have used \eqref{eq:Bessel_sincos} to replace Bessel functions of order $1/2$. In the matter dominated era we have $\alpha=2$, and so,
\begin{align}
    G_{\mathbf{k}}(\tau,\bar{\tau})=\dfrac{1}{k}\bigg[&\left(\dfrac{\bar{\tau}-\tau}{\tau\bar{\tau}}\right)\cos k(\tau-\bar{\tau}) +\nonumber\\
    &\left(\dfrac{1/k^2-\tau\bar{\tau}}{\tau\bar{\tau}}\right)\sin k(\tau-\bar{\tau})\bigg]\,.
\end{align}
where we have again used \eqref{eq:Bessel_sincos} to replace Bessel functions of order $3/2$.

Having calculated the Green's functions, we can now write the solution for $h_\lambda(\tau,\mathbf{k})$ in the form of \eqref{eq:GW_solution}.

\subsection{Connected and disconnected 4-point correlation function}\label{app:conn_disconn}
The primordial 4-point correlation function of $\zeta$ can be written in terms of disconnected and connected pieces
\begin{align}
    \langle\zeta(\mathbf{k}_1)\zeta(\mathbf{k}_2)\zeta(\mathbf{k}_3)\zeta(\mathbf{k}_4)\rangle=&\langle\zeta(\mathbf{k}_1)\zeta(\mathbf{k}_2)\zeta(\mathbf{k}_3)\zeta(\mathbf{k}_4)\rangle_\mathrm{d}\nonumber\\
    &+\langle\zeta(\mathbf{k}_1)\zeta(\mathbf{k}_2)\zeta(\mathbf{k}_3)\zeta(\mathbf{k}_4)\rangle_\mathrm{c}, 
\end{align}
where
\begin{align}
    &\langle\zeta(\mathbf{k}_1)\zeta(\mathbf{k}_2)\zeta(\mathbf{k}_3)\zeta(\mathbf{k}_4)\rangle_\mathrm{d}= \nonumber\\
    &(2\pi)^{6}\delta^3(\mathbf{k}_1+\mathbf{k}_2)\delta^3(\mathbf{k}_3+\mathbf{k}_4) P_\zeta(k_1) P_\zeta(k_3)\nonumber\\
    &+(2\pi)^{6}\delta^3(\mathbf{k}_1+\mathbf{k}_3)\delta^3(\mathbf{k}_2+\mathbf{k}_4) P_\zeta(k_1) P_\zeta(k_2)\nonumber\\
    &+(2\pi)^{6}\delta^3(\mathbf{k}_1+\mathbf{k}_4)\delta^3(\mathbf{k}_2+\mathbf{k}_4) P_\zeta(k_1) P_\zeta(k_4)\,,
\end{align}
and 
\begin{align}    &\langle\zeta(\mathbf{k}_1)\zeta(\mathbf{k}_2)\zeta(\mathbf{k}_3)\zeta(\mathbf{k}_4)\rangle_\mathrm{c} =\nonumber\\
    &(2\pi)^{3}\delta^3(\mathbf{k}_1+\mathbf{k}_2+\mathbf{k}_3+\mathbf{k}_4)\mathcal{T}(\mathbf{k}_1,\mathbf{k}_2,\mathbf{k}_3,\mathbf{k}_4)\,.
\end{align}
Here, $ P_\zeta(k)$ and $\mathcal{T}(\mathbf{k}_1,\mathbf{k}_2,\mathbf{k}_3,\mathbf{k}_4)$ are the scalar power spectrum and trispectrum, respectively. We focus on the disconnected contribution below. The relevant 4-point correlation function for the GW power spectrum \eqref{eq:master} is
\begin{align}
    &\langle\zeta(\mathbf{q}_1)\zeta(\mathbf{k}_1-\mathbf{q}_1)\zeta(\mathbf{q}_2)\zeta(\mathbf{k}_2-\mathbf{q}_2)\rangle_\mathrm{d}= \nonumber\\
    &~(2\pi)^{6}\delta^3(\mathbf{k}_1+\mathbf{k}_2)\left[\delta^3(\mathbf{q}_1+\mathbf{q}_2)+\delta^3(\mathbf{k}_1+\mathbf{q}_2-\mathbf{q}_1)\right]\nonumber\\
    &\times  P_\zeta(q_1) P_\zeta(|\mathbf{k}_1-\mathbf{q}_1|).
\end{align}
The two terms in the above expressions are equivalent when substituted in the integrand of \eqref{eq:master}. The second term can be manipulated as
\begin{align}
    &\delta^3(\mathbf{k}_1+\mathbf{k}_2)\delta^3(\mathbf{k}_1+\mathbf{q}_2-\mathbf{q}_1)Q_{\lambda_1}(\mathbf{k}_1,\mathbf{q}_1)Q_{\lambda_2}(\mathbf{k}_2,\mathbf{q}_2)\nonumber\\
    &~~~~~~~\times I(|\mathbf{k}_1-\mathbf{q}_1|,q_1,\tau)I(|\mathbf{k}_2-\mathbf{q}_2|,q_2,\tau) \nonumber\\
    & = Q_{\lambda_1}(\mathbf{k}_1,\mathbf{q}_1)Q_{\lambda_2}(-\mathbf{k}_1,\mathbf{q}_1-\mathbf{k}_1) I(|\mathbf{k}_1-\mathbf{q}_1|,q_1,\tau)\nonumber\\
    &~~~~~~~\times I(q_1,|\mathbf{k}_1-\mathbf{q}_1|,\tau) \nonumber\\
    & = Q_{\lambda_1}(\mathbf{k}_1,\mathbf{q}_1)^2I(|\mathbf{k}_1-\mathbf{q}_1|,q_1,\tau)^2
    \label{eq:discon_term2}
\end{align}
which is the same result we get from the first term. Here we have used identities given in eqs. \eqref{eq:Q_momentum_flip_symm}-\eqref{eq:I_exchange_symm}. Thus, the disconnected GW power spectrum is given by \eqref{eq:GW_power_disconnected}.

\subsection{Recasting integrals for numerical computation}\label{app:recast_numeric}
Here we provide steps to recast \eqref{eq:GW_power_disconnected} into a form suitable for numerical integration. \\
\paragraph{Change of variables.} We perform two successive changes of variables to recast the integrals. First, we perform the transformation $\{q,\cos\theta\}\rightarrow\{u,v\}$, where
\begin{equation}
    u\equiv\dfrac{|\mathbf{k}-\mathbf{q}|}{k}\,,\quad v\equiv\dfrac{q}{k},
\end{equation}
and the inverse transformation is 
\begin{equation}\label{eq:costheta}
    q=vk\,,\quad \cos\theta=\dfrac{1+v^2-u^2}{2v}\,.
\end{equation}
The determinant of the Jacobian for this transformation is,
\begin{equation}
    {\rm det}(J_{\{q,\cos\theta\}\rightarrow\{u,v\}})=-\partial_{v}q\partial_{u}\cos\theta=-\dfrac{ku}{v}\,.
\end{equation}
which implies
\begin{align}
    \int\mathrm{d}^3q&=\int_0^\infty q^2\mathrm{d} q \int_{-1}^{1}\mathrm{d}\cos\theta \int_0^{2\pi}\mathrm{d}\phi \nonumber\\
    & = k^3\int_0^\infty\mathrm{d} v\,v\int_{|1-v|}^{1+v}\mathrm{d} u\,u \int_0^{2\pi}\mathrm{d}\phi\,.
\end{align}
Second, we perform $\{u,v\}\rightarrow\{s,t\}$ where
\begin{equation}
    s\equiv u-v\,,\quad t \equiv u+v-1\,,
\end{equation}
and the inverse transformation is 
\begin{equation}\label{eq:uv_definition}
    u=\dfrac{s+t+1}{2}\,,\quad v=\dfrac{t-s+1}{2}\,.
\end{equation}
The determinant of the Jacobian for the second transformation is then
\begin{equation}
    {\rm det}(J_{\{u,v\}\rightarrow\{s,t\}})=\dfrac{1}{2}\,.
\end{equation}
Hence, we have\footnote{For $v<1$, the lower limit of integration over $s$ is $1-2v$. However, in this case we already have $1-2v>-1$.}
\begin{equation}
    \int_0^\infty\mathrm{d} v\int_{|1-v|}^{1+v}\mathrm{d} u = \dfrac{1}{2}\int_0^\infty\mathrm{d} t\int_{-1}^{1}\mathrm{d} s.
\end{equation}
The final result is 
\begin{equation}\label{eq:int_measure_change_of_var_FINAL}
    \int\mathrm{d}^3q=\dfrac{k^3}{2}\int_0^\infty\mathrm{d} t\int_{-1}^{1}\mathrm{d} s~ uv \int_0^{2\pi}\mathrm{d}\phi\,.
\end{equation}
Above, we express the integrand in terms of $u$ and $v$ for convenience, though the integration itself is done in terms of $s$ and $t$. \\

\paragraph{Analytic result for the \texorpdfstring{$I(p,q,\tau)$}{I(p,q,tau)} function.}
We summarize the results for a radiation-dominated universe (for a more in-depth look, see e.g. \cite{Kohri:2018awv}). At late times, we have
\begin{align}
    I(vk,uk,x/k\rightarrow\infty)=\dfrac{1}{k^2}I(u,v,x\rightarrow\infty)\nonumber\\
    \simeq \dfrac{1}{k^2}\dfrac{1}{x}\tilde{I}_A(u,v)\left(\tilde{I}_B(u,v)\sin x + \tilde{I}_C\cos x\right),
\end{align}
where we define
\begin{subequations}\label{eq:Itilde}
\begin{align}
    \tilde{I}_A(u,v) &\equiv \dfrac{3(u^2+v^2-3)}{4u^3v^3}\\
    \tilde{I}_B(u,v) &\equiv -4uv+(u^2+v^2-3)\ln\left|\dfrac{3-(u+v)^2}{3-(u-v)^2}\right|\\
    \tilde{I}_C(u,v) &\equiv -\pi(u^2+v^2-3)\Theta(u+v-\sqrt{3})\,.
\end{align}
\end{subequations}
In the last expression, $\Theta$ is the Heaviside theta function. This result redshifts as $1/x\propto1/a$. Using the above definitions, we compute the quantity given in \ref{eq:discon_term2},
\begin{align}
    &\dfrac{Q_{+}(\mathbf{k},\mathbf{q})}{\cos2\phi}I(|\mathbf{k}-\mathbf{q}|,q,\tau)\nonumber\\
    &~=\dfrac{Q_{\times}(\mathbf{k},\mathbf{q})}{\sin2\phi}I(|\mathbf{k}-\mathbf{q}|,q,\tau)\nonumber\\
    &~=\dfrac{v^2k^2}{\sqrt{2}}\dfrac{4v^2-(1+v^2-u^2)^2}{4v^2}I(uk,vk,x/k)\nonumber\\
    &~\equiv\dfrac{\Tilde{\mathcal{J}}(u,v)}{\sqrt{2}}k^2I(uk,vk,x/k),
\end{align}
where we have used dimensionless conformal time $x=k\tau$ and defined
\begin{equation}\label{eq:Jtilde}
    \Tilde{\mathcal{J}}(u,v) = \dfrac{4v^2-(1+v^2-u^2)^2}{4}\,.
\end{equation}
When computing the GW power spectrum we are generically interested in the time-averaged quantity
\begin{align}
    &\overline{k^2I(v_1k,u_1k,x/k\rightarrow\infty)k^2I(v_2k,u_2k,x/k\rightarrow\infty)} = \nonumber\\
    ~&\dfrac{1}{2x^2}\tilde{I}_A(u_1,v_1) \tilde{I}_A(u_2,v_2)\nonumber\\
    &\times\Big[\tilde{I}_B(u_1,v_1) \tilde{I}_B(u_2,v_2)+\tilde{I}_C(u_1,v_1) \tilde{I}_C(u_2,v_2)\Big].
\end{align} 
\paragraph{Azimuthal angle integration.} In the  disconnected contribution \eqref{eq:GW_power_disconnected}, the only $\phi$-dependent factors in the integrands are $\sin2\phi$ and $\cos2\phi$, coming from $Q_\lambda$ factors. For each polarization, we then have
\begin{equation}\label{eq:azimuthal_disconnected}
    \int_0^{2\pi}\mathrm{d}\phi \sin^2(2\phi)=\int_0^{2\pi}\mathrm{d}\phi \cos^2(2\phi)=\pi\,.
\end{equation}

Finally, we are ready to numerically compute the GW energy density \eqref{eq:GW_energy_density} which is defined in terms of the dimensionless polarization-averaged GW power spectrum
\begin{equation}
    \sum_\lambda \Delta_\lambda^2(\tau,k) = \dfrac{k^3}{2\pi^2} \sum_\lambda  P_\lambda(\tau,k).
\end{equation}
Using our recasted variables, the result is
\begin{align}\label{eq:GW_power_disconnected_recasted}
    \Omega_{\text{\tiny{GW}}}(k)\bigg|_\mathrm{d} = &\dfrac{2}{48\alpha^2}\left(\dfrac{k^3}{2\pi^2}\right)^2\nonumber\\
    &\int_0^\infty\mathrm{d} t\int_{-1}^{1}\mathrm{d} s uv \Tilde{\mathcal{J}}(u,v)^2\tilde{I}_A(u,v)^2\bigg[\tilde{I}_B(u,v)^2\nonumber\\
    &\quad \quad+\tilde{I}_C(u,v)^2\bigg] P_\zeta(uk) P_\zeta(vk)
\end{align}

More compactly,
\begin{align}
    \Omega_{\text{\tiny{GW}}}(k)\bigg|_\mathrm{d} = \dfrac{2}{48\alpha^2}\int_0^\infty\mathrm{d} t\int_{-1}^{1}\mathrm{d} s ~\mathcal{K}_{\mathrm{d}}(u,v)  \Delta^2_\zeta(uk) \Delta^2_\zeta(vk) \label{eq:disconnected_general_final}
\end{align}
where we define the following the Kernel functions $\mathcal{K}_{\mathrm{d}}$ for simplified notation,
\begin{equation}
    \mathcal{K}_{\mathrm{d}}(u,v) = (uv)^{-2}\Tilde{\mathcal{J}}(u,v)^2\tilde{I}_A(u,v)^2\left[\tilde{I}_B(u,v)^2+\tilde{I}_C(u,v)^2\right].\label{eq:Kcal_disconnected}
\end{equation}

\subsection{Useful formula}
The projection operator $Q_\lambda$ \eqref{eq:Q_definition} is defined as,

\begin{align}
    Q_\lambda(\mathbf{k},\mathbf{q})\equiv\epsilon_\lambda^{ij}(\mathbf{k})q_iq_j=-\epsilon_\lambda^{ij}(\mathbf{k})(\mathbf{k}-\mathbf{q})_iq_j,
\end{align}

where the second equality follows from $\epsilon_\lambda^{ij}(\mathbf{k})k_i=0$. If we explicitly set $\hat{k}=\hat{z}$, we have $\mathbf{q}=q(\sin\theta\cos\phi, \sin\theta\sin\phi, \cos\theta)$, where $\theta$ and $\phi$ are polar and azimuthal angles. This leads to the expressions,
\begin{align}\label{eq:Q_factors_explicit}
    &Q_{+}(\mathbf{k},\mathbf{q}) = \dfrac{q^2}{\sqrt{2}}\sin^2\theta\cos(2\phi)\,,\nonumber\\
    &Q_{\times}(\mathbf{k},\mathbf{q}) = \dfrac{q^2}{\sqrt{2}}\sin^2\theta\sin(2\phi)\,.
\end{align}
Since $\epsilon_\lambda(\mathbf{k})$ is orthogonal to $\mathbf{k}$ we have
\begin{align}\label{eq:Q_orhtogonal_relation}
    Q_\lambda(\mathbf{k},\mathbf{q})=Q_\lambda(\mathbf{k},\mathbf{q}+c\mathbf{k})\,,
\end{align}
for any constant $c$. $Q_\lambda(\mathbf{k},\mathbf{q})$ is also symmetric under $\mathbf{k}\rightarrow-\mathbf{k}$ and $\mathbf{q}\rightarrow-\mathbf{q}$:
\begin{equation}\label{eq:Q_momentum_flip_symm}
    Q_\lambda(\mathbf{k},\mathbf{q})=Q_\lambda(-\mathbf{k},\mathbf{q})=Q_\lambda(\mathbf{k},-\mathbf{q})=Q_\lambda(-\mathbf{k},-\mathbf{q})\,.
\end{equation}
Using \eqref{eq:f_function} we see that
\begin{equation}
    f(p,q,\tau)=f(q,p,\tau)
\end{equation}
and so
\begin{equation}\label{eq:I_exchange_symm}
     I(p,q,\tau)=I(q,p,\tau)\,.
\end{equation}

\paragraph{Bessel functions.} The following formulae are helpful for computations involving Bessel functions:
\begin{align}\label{eq:Bessel_sincos}
    J_{1/2}(x)&=\sqrt{\dfrac{2}{\pi x}}\sin x\,,\nonumber\\
    Y_{1/2}(x) &=-\sqrt{\dfrac{2}{\pi x}}\cos x\,,\nonumber\\
    J_{3/2}(x)&=\sqrt{\dfrac{2}{\pi x}}\left(\dfrac{\sin x}{x}-\cos x\right)\,,\nonumber\\
    Y_{3/2}(x) &=-\sqrt{\dfrac{2}{\pi x}}\left(\dfrac{\cos x}{x}-\sin x\right)\,.
\end{align}

\bibliographystyle{utphys}
\bibliography{references}

\providecommand{\href}[2]{#2}\begingroup\raggedright\begin{thebibliography}{10}

\bibitem{Baumann:2009ds}
D.~Baumann,
  \href{http://dx.doi.org/10.1142/9789814327183_0010}{``{Inflation},''} in {\em
  {Theoretical Advanced Study Institute in Elementary Particle Physics}:
  {Physics of the Large and the Small}}, pp.~523--686.
\newblock 2011.
\newblock \href{http://arxiv.org/abs/0907.5424}{{\ttfamily arXiv:0907.5424
  [hep-th]}}.

\bibitem{Planck:2018jri}
{\bfseries Planck} Collaboration, Y.~Akrami {\em et~al.}, ``{Planck 2018
  results. X. Constraints on inflation},''
  \href{http://dx.doi.org/10.1051/0004-6361/201833887}{{\em Astron. Astrophys.}
  {\bfseries 641} (2020) A10},
  \href{http://arxiv.org/abs/1807.06211}{{\ttfamily arXiv:1807.06211
  [astro-ph.CO]}}.

\bibitem{Boddy:2022knd}
K.~K. Boddy {\em et~al.}, ``{Snowmass2021 theory frontier white paper:
  Astrophysical and cosmological probes of dark matter},''
  \href{http://dx.doi.org/10.1016/j.jheap.2022.06.005}{{\em JHEAp} {\bfseries
  35} (2022) 112--138}, \href{http://arxiv.org/abs/2203.06380}{{\ttfamily
  arXiv:2203.06380 [hep-ph]}}.

\bibitem{Domenech:2021ztg}
G.~Dom\`enech, ``{Scalar Induced Gravitational Waves Review},''
  \href{http://dx.doi.org/10.3390/universe7110398}{{\em Universe} {\bfseries 7}
  no.~11, (2021) 398}, \href{http://arxiv.org/abs/2109.01398}{{\ttfamily
  arXiv:2109.01398 [gr-qc]}}.

\bibitem{Green:2020jor}
A.~M. Green and B.~J. Kavanagh, ``{Primordial Black Holes as a dark matter
  candidate},'' \href{http://dx.doi.org/10.1088/1361-6471/abc534}{{\em J. Phys.
  G} {\bfseries 48} no.~4, (2021) 043001},
  \href{http://arxiv.org/abs/2007.10722}{{\ttfamily arXiv:2007.10722
  [astro-ph.CO]}}.

\bibitem{Carr:2020xqk}
B.~Carr and F.~Kuhnel, ``{Primordial Black Holes as Dark Matter: Recent
  Developments},''
  \href{http://dx.doi.org/10.1146/annurev-nucl-050520-125911}{{\em Ann. Rev.
  Nucl. Part. Sci.} {\bfseries 70} (2020) 355--394},
  \href{http://arxiv.org/abs/2006.02838}{{\ttfamily arXiv:2006.02838
  [astro-ph.CO]}}.

\bibitem{Ivanov:1994pa}
P.~Ivanov, P.~Naselsky, and I.~Novikov, ``{Inflation and primordial black holes
  as dark matter},'' \href{http://dx.doi.org/10.1103/PhysRevD.50.7173}{{\em
  Phys. Rev. D} {\bfseries 50} (1994) 7173--7178}.

\bibitem{Garcia-Bellido:2017mdw}
J.~Garcia-Bellido and E.~Ruiz~Morales, ``{Primordial black holes from single
  field models of inflation},''
  \href{http://dx.doi.org/10.1016/j.dark.2017.09.007}{{\em Phys. Dark Univ.}
  {\bfseries 18} (2017) 47--54},
  \href{http://arxiv.org/abs/1702.03901}{{\ttfamily arXiv:1702.03901
  [astro-ph.CO]}}.

\bibitem{Ballesteros:2017fsr}
G.~Ballesteros and M.~Taoso, ``{Primordial black hole dark matter from single
  field inflation},'' \href{http://dx.doi.org/10.1103/PhysRevD.97.023501}{{\em
  Phys. Rev. D} {\bfseries 97} no.~2, (2018) 023501},
  \href{http://arxiv.org/abs/1709.05565}{{\ttfamily arXiv:1709.05565
  [hep-ph]}}.

\bibitem{Tsamis:2003px}
N.~C. Tsamis and R.~P. Woodard, ``{Improved estimates of cosmological
  perturbations},'' \href{http://dx.doi.org/10.1103/PhysRevD.69.084005}{{\em
  Phys. Rev. D} {\bfseries 69} (2004) 084005},
  \href{http://arxiv.org/abs/astro-ph/0307463}{{\ttfamily
  arXiv:astro-ph/0307463}}.

\bibitem{Kinney:2005vj}
W.~H. Kinney, ``{Horizon crossing and inflation with large eta},''
  \href{http://dx.doi.org/10.1103/PhysRevD.72.023515}{{\em Phys. Rev. D}
  {\bfseries 72} (2005) 023515},
  \href{http://arxiv.org/abs/gr-qc/0503017}{{\ttfamily arXiv:gr-qc/0503017}}.

\bibitem{Hooshangi:2022lao}
S.~Hooshangi, A.~Talebian, M.~H. Namjoo, and H.~Firouzjahi, ``{Multiple field
  ultraslow-roll inflation: Primordial black holes from straight bulk and
  distorted boundary},''
  \href{http://dx.doi.org/10.1103/PhysRevD.105.083525}{{\em Phys. Rev. D}
  {\bfseries 105} no.~8, (2022) 083525},
  \href{http://arxiv.org/abs/2201.07258}{{\ttfamily arXiv:2201.07258
  [astro-ph.CO]}}.

\bibitem{Kasuya:2009up}
S.~Kasuya and M.~Kawasaki, ``{Axion isocurvature fluctuations with extremely
  blue spectrum},'' \href{http://dx.doi.org/10.1103/PhysRevD.80.023516}{{\em
  Phys. Rev. D} {\bfseries 80} (2009) 023516},
  \href{http://arxiv.org/abs/0904.3800}{{\ttfamily arXiv:0904.3800
  [astro-ph.CO]}}.

\bibitem{Kawasaki:2012wr}
M.~Kawasaki, N.~Kitajima, and T.~T. Yanagida, ``{Primordial black hole
  formation from an axionlike curvaton model},''
  \href{http://dx.doi.org/10.1103/PhysRevD.87.063519}{{\em Phys. Rev. D}
  {\bfseries 87} no.~6, (2013) 063519},
  \href{http://arxiv.org/abs/1207.2550}{{\ttfamily arXiv:1207.2550 [hep-ph]}}.

\bibitem{Chung:2015pga}
D.~J.~H. Chung and H.~Yoo, ``{Elementary Theorems Regarding Blue Isocurvature
  Perturbations},'' \href{http://dx.doi.org/10.1103/PhysRevD.91.083530}{{\em
  Phys. Rev. D} {\bfseries 91} (2015) 083530},
  \href{http://arxiv.org/abs/1501.05618}{{\ttfamily arXiv:1501.05618
  [astro-ph.CO]}}.

\bibitem{Chung:2017uzc}
D.~J.~H. Chung and A.~Upadhye, ``{Search for strongly blue axion
  isocurvature},'' \href{http://dx.doi.org/10.1103/PhysRevD.98.023525}{{\em
  Phys. Rev. D} {\bfseries 98} no.~2, (2018) 023525},
  \href{http://arxiv.org/abs/1711.06736}{{\ttfamily arXiv:1711.06736
  [astro-ph.CO]}}.

\bibitem{Chung:2021lfg}
D.~J.~H. Chung and S.~C. Tadepalli, ``{Analytic treatment of underdamped
  axionic blue isocurvature perturbations},''
  \href{http://dx.doi.org/10.1103/PhysRevD.105.123511}{{\em Phys. Rev. D}
  {\bfseries 105} no.~12, (2022) 123511},
  \href{http://arxiv.org/abs/2110.02272}{{\ttfamily arXiv:2110.02272
  [astro-ph.CO]}}.

\bibitem{Talebian:2022jkb}
A.~Talebian, A.~Nassiri-Rad, and H.~Firouzjahi, ``{Stochastic effects in axion
  inflation and primordial black hole formation},''
  \href{http://dx.doi.org/10.1103/PhysRevD.105.103516}{{\em Phys. Rev. D}
  {\bfseries 105} no.~10, (2022) 103516},
  \href{http://arxiv.org/abs/2202.02062}{{\ttfamily arXiv:2202.02062
  [astro-ph.CO]}}.

\bibitem{Graham:2015rva}
P.~W. Graham, J.~Mardon, and S.~Rajendran, ``{Vector Dark Matter from
  Inflationary Fluctuations},''
  \href{http://dx.doi.org/10.1103/PhysRevD.93.103520}{{\em Phys. Rev. D}
  {\bfseries 93} no.~10, (2016) 103520},
  \href{http://arxiv.org/abs/1504.02102}{{\ttfamily arXiv:1504.02102
  [hep-ph]}}.

\bibitem{Erickcek:2011us}
A.~L. Erickcek and K.~Sigurdson, ``{Reheating Effects in the Matter Power
  Spectrum and Implications for Substructure},''
  \href{http://dx.doi.org/10.1103/PhysRevD.84.083503}{{\em Phys. Rev. D}
  {\bfseries 84} (2011) 083503},
  \href{http://arxiv.org/abs/1106.0536}{{\ttfamily arXiv:1106.0536
  [astro-ph.CO]}}.

\bibitem{Barir:2022kzo}
J.~Barir, M.~Geller, C.~Sun, and T.~Volansky, ``{Gravitational Waves from
  Incomplete Inflationary Phase Transitions},''
  \href{http://arxiv.org/abs/2203.00693}{{\ttfamily arXiv:2203.00693
  [hep-ph]}}.

\bibitem{Chung:2004nh}
D.~J.~H. Chung, E.~W. Kolb, A.~Riotto, and L.~Senatore, ``{Isocurvature
  constraints on gravitationally produced superheavy dark matter},''
  \href{http://dx.doi.org/10.1103/PhysRevD.72.023511}{{\em Phys. Rev. D}
  {\bfseries 72} (2005) 023511},
  \href{http://arxiv.org/abs/astro-ph/0411468}{{\ttfamily
  arXiv:astro-ph/0411468}}.

\bibitem{Starobinsky:1986fx}
A.~A. Starobinsky, ``{STOCHASTIC DE SITTER (INFLATIONARY) STAGE IN THE EARLY
  UNIVERSE},'' \href{http://dx.doi.org/10.1007/3-540-16452-9_6}{{\em Lect.
  Notes Phys.} {\bfseries 246} (1986) 107--126}.

\bibitem{Starobinsky:1994bd}
A.~A. Starobinsky and J.~Yokoyama, ``{Equilibrium state of a selfinteracting
  scalar field in the De Sitter background},''
  \href{http://dx.doi.org/10.1103/PhysRevD.50.6357}{{\em Phys. Rev. D}
  {\bfseries 50} (1994) 6357--6368},
  \href{http://arxiv.org/abs/astro-ph/9407016}{{\ttfamily
  arXiv:astro-ph/9407016}}.

\bibitem{Linde:1996gt}
A.~D. Linde and V.~F. Mukhanov, ``{Nongaussian isocurvature perturbations from
  inflation},'' \href{http://dx.doi.org/10.1103/PhysRevD.56.R535}{{\em Phys.
  Rev. D} {\bfseries 56} (1997) R535--R539},
  \href{http://arxiv.org/abs/astro-ph/9610219}{{\ttfamily
  arXiv:astro-ph/9610219}}.

\bibitem{Enqvist:2001zp}
K.~Enqvist and M.~S. Sloth, ``{Adiabatic CMB perturbations in pre - big bang
  string cosmology},''
  \href{http://dx.doi.org/10.1016/S0550-3213(02)00043-3}{{\em Nucl. Phys. B}
  {\bfseries 626} (2002) 395--409},
  \href{http://arxiv.org/abs/hep-ph/0109214}{{\ttfamily arXiv:hep-ph/0109214}}.

\bibitem{Moroi:2001ct}
T.~Moroi and T.~Takahashi, ``{Effects of cosmological moduli fields on cosmic
  microwave background},''
  \href{http://dx.doi.org/10.1016/S0370-2693(01)01295-3}{{\em Phys. Lett. B}
  {\bfseries 522} (2001) 215--221},
  \href{http://arxiv.org/abs/hep-ph/0110096}{{\ttfamily arXiv:hep-ph/0110096}}.
  [Erratum: Phys.Lett.B 539, 303--303 (2002)].

\bibitem{Lyth:2001nq}
D.~H. Lyth and D.~Wands, ``{Generating the curvature perturbation without an
  inflaton},'' \href{http://dx.doi.org/10.1016/S0370-2693(01)01366-1}{{\em
  Phys. Lett. B} {\bfseries 524} (2002) 5--14},
  \href{http://arxiv.org/abs/hep-ph/0110002}{{\ttfamily arXiv:hep-ph/0110002}}.

\bibitem{Kolb:1990vq}
E.~W. Kolb and M.~S. Turner,
  \href{http://dx.doi.org/10.1201/9780429492860}{{\em {The Early Universe}}},
  vol.~69.
\newblock 1990.

\bibitem{Malik:2008im}
K.~A. Malik and D.~Wands, ``{Cosmological perturbations},''
  \href{http://dx.doi.org/10.1016/j.physrep.2009.03.001}{{\em Phys. Rept.}
  {\bfseries 475} (2009) 1--51},
  \href{http://arxiv.org/abs/0809.4944}{{\ttfamily arXiv:0809.4944
  [astro-ph]}}.

\bibitem{Wands:2000dp}
D.~Wands, K.~A. Malik, D.~H. Lyth, and A.~R. Liddle, ``{A New approach to the
  evolution of cosmological perturbations on large scales},''
  \href{http://dx.doi.org/10.1103/PhysRevD.62.043527}{{\em Phys. Rev. D}
  {\bfseries 62} (2000) 043527},
  \href{http://arxiv.org/abs/astro-ph/0003278}{{\ttfamily
  arXiv:astro-ph/0003278}}.

\bibitem{Sasaki:1987gy}
M.~Sasaki, Y.~Nambu, and K.-i. Nakao, ``{Classical Behavior of a Scalar Field
  in the Inflationary Universe},''
  \href{http://dx.doi.org/10.1016/0550-3213(88)90132-0}{{\em Nucl. Phys. B}
  {\bfseries 308} (1988) 868--884}.

\bibitem{Nambu:1987ef}
Y.~Nambu and M.~Sasaki, ``{Stochastic Stage of an Inflationary Universe
  Model},'' \href{http://dx.doi.org/10.1016/0370-2693(88)90974-4}{{\em Phys.
  Lett. B} {\bfseries 205} (1988) 441--446}.

\bibitem{Graham:2018jyp}
P.~W. Graham and A.~Scherlis, ``{Stochastic axion scenario},''
  \href{http://dx.doi.org/10.1103/PhysRevD.98.035017}{{\em Phys. Rev. D}
  {\bfseries 98} no.~3, (2018) 035017},
  \href{http://arxiv.org/abs/1805.07362}{{\ttfamily arXiv:1805.07362
  [hep-ph]}}.

\bibitem{Markkanen:2019kpv}
T.~Markkanen, A.~Rajantie, S.~Stopyra, and T.~Tenkanen, ``{Scalar correlation
  functions in de Sitter space from the stochastic spectral expansion},''
  \href{http://dx.doi.org/10.1088/1475-7516/2019/08/001}{{\em JCAP} {\bfseries
  08} (2019) 001}, \href{http://arxiv.org/abs/1904.11917}{{\ttfamily
  arXiv:1904.11917 [gr-qc]}}.

\bibitem{Liddle:2003as}
A.~R. Liddle and S.~M. Leach, ``{How long before the end of inflation were
  observable perturbations produced?},''
  \href{http://dx.doi.org/10.1103/PhysRevD.68.103503}{{\em Phys. Rev. D}
  {\bfseries 68} (2003) 103503},
  \href{http://arxiv.org/abs/astro-ph/0305263}{{\ttfamily
  arXiv:astro-ph/0305263}}.

\bibitem{Dodelson:2003vq}
S.~Dodelson and L.~Hui, ``{A Horizon ratio bound for inflationary
  fluctuations},'' \href{http://dx.doi.org/10.1103/PhysRevLett.91.131301}{{\em
  Phys. Rev. Lett.} {\bfseries 91} (2003) 131301},
  \href{http://arxiv.org/abs/astro-ph/0305113}{{\ttfamily
  arXiv:astro-ph/0305113}}.

\bibitem{Abbott:1982hn}
L.~F. Abbott, E.~Farhi, and M.~B. Wise, ``{Particle Production in the New
  Inflationary Cosmology},''
  \href{http://dx.doi.org/10.1016/0370-2693(82)90867-X}{{\em Phys. Lett. B}
  {\bfseries 117} (1982) 29}.

\bibitem{Dolgov:1982th}
A.~D. Dolgov and A.~D. Linde, ``{Baryon Asymmetry in Inflationary Universe},''
  \href{http://dx.doi.org/10.1016/0370-2693(82)90292-1}{{\em Phys. Lett. B}
  {\bfseries 116} (1982) 329}.

\bibitem{Albrecht:1982mp}
A.~Albrecht, P.~J. Steinhardt, M.~S. Turner, and F.~Wilczek, ``{Reheating an
  Inflationary Universe},''
  \href{http://dx.doi.org/10.1103/PhysRevLett.48.1437}{{\em Phys. Rev. Lett.}
  {\bfseries 48} (1982) 1437}.

\bibitem{Podolsky:2005bw}
D.~I. Podolsky, G.~N. Felder, L.~Kofman, and M.~Peloso, ``{Equation of state
  and beginning of thermalization after preheating},''
  \href{http://dx.doi.org/10.1103/PhysRevD.73.023501}{{\em Phys. Rev. D}
  {\bfseries 73} (2006) 023501},
  \href{http://arxiv.org/abs/hep-ph/0507096}{{\ttfamily arXiv:hep-ph/0507096}}.

\bibitem{Munoz:2014eqa}
J.~B. Munoz and M.~Kamionkowski, ``{Equation-of-State Parameter for
  Reheating},'' \href{http://dx.doi.org/10.1103/PhysRevD.91.043521}{{\em Phys.
  Rev. D} {\bfseries 91} no.~4, (2015) 043521},
  \href{http://arxiv.org/abs/1412.0656}{{\ttfamily arXiv:1412.0656
  [astro-ph.CO]}}.

\bibitem{Lozanov:2016hid}
K.~D. Lozanov and M.~A. Amin, ``{Equation of State and Duration to Radiation
  Domination after Inflation},''
  \href{http://dx.doi.org/10.1103/PhysRevLett.119.061301}{{\em Phys. Rev.
  Lett.} {\bfseries 119} no.~6, (2017) 061301},
  \href{http://arxiv.org/abs/1608.01213}{{\ttfamily arXiv:1608.01213
  [astro-ph.CO]}}.

\bibitem{Maity:2018qhi}
D.~Maity and P.~Saha, ``{(P)reheating after minimal Plateau Inflation and
  constraints from CMB},''
  \href{http://dx.doi.org/10.1088/1475-7516/2019/07/018}{{\em JCAP} {\bfseries
  07} (2019) 018}, \href{http://arxiv.org/abs/1811.11173}{{\ttfamily
  arXiv:1811.11173 [astro-ph.CO]}}.

\bibitem{Antusch:2020iyq}
S.~Antusch, D.~G. Figueroa, K.~Marschall, and F.~Torrenti, ``{Energy
  distribution and equation of state of the early Universe: matching the end of
  inflation and the onset of radiation domination},''
  \href{http://dx.doi.org/10.1016/j.physletb.2020.135888}{{\em Phys. Lett. B}
  {\bfseries 811} (2020) 135888},
  \href{http://arxiv.org/abs/2005.07563}{{\ttfamily arXiv:2005.07563
  [astro-ph.CO]}}.

\bibitem{Allahverdi:2010xz}
R.~Allahverdi, R.~Brandenberger, F.-Y. Cyr-Racine, and A.~Mazumdar,
  ``{Reheating in Inflationary Cosmology: Theory and Applications},''
  \href{http://dx.doi.org/10.1146/annurev.nucl.012809.104511}{{\em Ann. Rev.
  Nucl. Part. Sci.} {\bfseries 60} (2010) 27--51},
  \href{http://arxiv.org/abs/1001.2600}{{\ttfamily arXiv:1001.2600 [hep-th]}}.

\bibitem{Chluba:2012gq}
J.~Chluba, R.~Khatri, and R.~A. Sunyaev, ``{CMB at 2x2 order: The dissipation
  of primordial acoustic waves and the observable part of the associated energy
  release},'' \href{http://dx.doi.org/10.1111/j.1365-2966.2012.21474.x}{{\em
  Mon. Not. Roy. Astron. Soc.} {\bfseries 425} (2012) 1129--1169},
  \href{http://arxiv.org/abs/1202.0057}{{\ttfamily arXiv:1202.0057
  [astro-ph.CO]}}.

\bibitem{Chluba:2019kpb}
J.~Chluba {\em et~al.}, ``{Spectral Distortions of the CMB as a Probe of
  Inflation, Recombination, Structure Formation and Particle Physics}:
  {Astro2020 Science White Paper},'' {\em Bull. Am. Astron. Soc.} {\bfseries
  51} no.~3, (2019) 184, \href{http://arxiv.org/abs/1903.04218}{{\ttfamily
  arXiv:1903.04218 [astro-ph.CO]}}.

\bibitem{Lee:2020wfn}
V.~S.~H. Lee, A.~Mitridate, T.~Trickle, and K.~M. Zurek, ``{Probing Small-Scale
  Power Spectra with Pulsar Timing Arrays},''
  \href{http://dx.doi.org/10.1007/JHEP06(2021)028}{{\em JHEP} {\bfseries 06}
  (2021) 028}, \href{http://arxiv.org/abs/2012.09857}{{\ttfamily
  arXiv:2012.09857 [astro-ph.CO]}}.

\bibitem{VanTilburg:2018ykj}
K.~Van~Tilburg, A.-M. Taki, and N.~Weiner, ``{Halometry from Astrometry},''
  \href{http://dx.doi.org/10.1088/1475-7516/2018/07/041}{{\em JCAP} {\bfseries
  07} (2018) 041}, \href{http://arxiv.org/abs/1804.01991}{{\ttfamily
  arXiv:1804.01991 [astro-ph.CO]}}.

\bibitem{2002ApJ...581..817F}
D.~J. {Fixsen} and J.~C. {Mather}, ``{The Spectral Results of the Far-Infrared
  Absolute Spectrophotometer Instrument on COBE},''
  \href{http://dx.doi.org/10.1086/344402}{{\em \apj} {\bfseries 581} no.~2,
  (Dec., 2002) 817--822}.

\bibitem{Ananda:2006af}
K.~N. Ananda, C.~Clarkson, and D.~Wands, ``{The Cosmological gravitational wave
  background from primordial density perturbations},''
  \href{http://dx.doi.org/10.1103/PhysRevD.75.123518}{{\em Phys. Rev. D}
  {\bfseries 75} (2007) 123518},
  \href{http://arxiv.org/abs/gr-qc/0612013}{{\ttfamily arXiv:gr-qc/0612013}}.

\bibitem{Baumann:2007zm}
D.~Baumann, P.~J. Steinhardt, K.~Takahashi, and K.~Ichiki, ``{Gravitational
  Wave Spectrum Induced by Primordial Scalar Perturbations},''
  \href{http://dx.doi.org/10.1103/PhysRevD.76.084019}{{\em Phys. Rev. D}
  {\bfseries 76} (2007) 084019},
  \href{http://arxiv.org/abs/hep-th/0703290}{{\ttfamily arXiv:hep-th/0703290}}.

\bibitem{Garcia-Saenz:2022tzu}
S.~Garcia-Saenz, L.~Pinol, S.~Renaux-Petel, and D.~Werth, ``{No-go theorem for
  scalar-trispectrum-induced gravitational waves},''
  \href{http://dx.doi.org/10.1088/1475-7516/2023/03/057}{{\em JCAP} {\bfseries
  03} (2023) 057}, \href{http://arxiv.org/abs/2207.14267}{{\ttfamily
  arXiv:2207.14267 [astro-ph.CO]}}.

\bibitem{Adshead:2021hnm}
P.~Adshead, K.~D. Lozanov, and Z.~J. Weiner, ``{Non-Gaussianity and the induced
  gravitational wave background},''
  \href{http://dx.doi.org/10.1088/1475-7516/2021/10/080}{{\em JCAP} {\bfseries
  10} (2021) 080}, \href{http://arxiv.org/abs/2105.01659}{{\ttfamily
  arXiv:2105.01659 [astro-ph.CO]}}.

\bibitem{Unal:2018yaa}
C.~Unal, ``{Imprints of Primordial Non-Gaussianity on Gravitational Wave
  Spectrum},'' \href{http://dx.doi.org/10.1103/PhysRevD.99.041301}{{\em Phys.
  Rev. D} {\bfseries 99} no.~4, (2019) 041301},
  \href{http://arxiv.org/abs/1811.09151}{{\ttfamily arXiv:1811.09151
  [astro-ph.CO]}}.

\bibitem{Atal:2021jyo}
V.~Atal and G.~Dom\`enech, ``{Probing non-Gaussianities with the high frequency
  tail of induced gravitational waves},''
  \href{http://dx.doi.org/10.1088/1475-7516/2021/06/001}{{\em JCAP} {\bfseries
  06} (2021) 001}, \href{http://arxiv.org/abs/2103.01056}{{\ttfamily
  arXiv:2103.01056 [astro-ph.CO]}}.

\bibitem{Maggiore:1999vm}
M.~Maggiore, ``{Gravitational wave experiments and early universe cosmology},''
  \href{http://dx.doi.org/10.1016/S0370-1573(99)00102-7}{{\em Phys. Rept.}
  {\bfseries 331} (2000) 283--367},
  \href{http://arxiv.org/abs/gr-qc/9909001}{{\ttfamily arXiv:gr-qc/9909001}}.

\bibitem{Kohri:2018awv}
K.~Kohri and T.~Terada, ``{Semianalytic calculation of gravitational wave
  spectrum nonlinearly induced from primordial curvature perturbations},''
  \href{http://dx.doi.org/10.1103/PhysRevD.97.123532}{{\em Phys. Rev. D}
  {\bfseries 97} no.~12, (2018) 123532},
  \href{http://arxiv.org/abs/1804.08577}{{\ttfamily arXiv:1804.08577 [gr-qc]}}.

\bibitem{Schmitz:2020syl}
K.~Schmitz, ``{New Sensitivity Curves for Gravitational-Wave Signals from
  Cosmological Phase Transitions},''
  \href{http://dx.doi.org/10.1007/JHEP01(2021)097}{{\em JHEP} {\bfseries 01}
  (2021) 097}, \href{http://arxiv.org/abs/2002.04615}{{\ttfamily
  arXiv:2002.04615 [hep-ph]}}.

\bibitem{Sesana:2019vho}
A.~Sesana {\em et~al.}, ``{Unveiling the gravitational universe at $\mu$-Hz
  frequencies},'' \href{http://dx.doi.org/10.1007/s10686-021-09709-9}{{\em
  Exper. Astron.} {\bfseries 51} no.~3, (2021) 1333--1383},
  \href{http://arxiv.org/abs/1908.11391}{{\ttfamily arXiv:1908.11391
  [astro-ph.IM]}}.

\bibitem{Braglia:2021fxn}
M.~Braglia and S.~Kuroyanagi, ``{Probing prerecombination physics by the
  cross-correlation of stochastic gravitational waves and CMB anisotropies},''
  \href{http://dx.doi.org/10.1103/PhysRevD.104.123547}{{\em Phys. Rev. D}
  {\bfseries 104} no.~12, (2021) 123547},
  \href{http://arxiv.org/abs/2106.03786}{{\ttfamily arXiv:2106.03786
  [astro-ph.CO]}}.

\bibitem{NANOGrav:2023gor}
{\bfseries NANOGrav} Collaboration, G.~Agazie {\em et~al.}, ``{The NANOGrav 15
  yr Data Set: Evidence for a Gravitational-wave Background},''
  \href{http://dx.doi.org/10.3847/2041-8213/acdac6}{{\em Astrophys. J. Lett.}
  {\bfseries 951} no.~1, (2023) L8},
  \href{http://arxiv.org/abs/2306.16213}{{\ttfamily arXiv:2306.16213
  [astro-ph.HE]}}.

\bibitem{NANOGrav:2023hvm}
{\bfseries NANOGrav} Collaboration, A.~Afzal {\em et~al.}, ``{The NANOGrav 15
  yr Data Set: Search for Signals from New Physics},''
  \href{http://dx.doi.org/10.3847/2041-8213/acdc91}{{\em Astrophys. J. Lett.}
  {\bfseries 951} no.~1, (2023) L11},
  \href{http://arxiv.org/abs/2306.16219}{{\ttfamily arXiv:2306.16219
  [astro-ph.HE]}}.

\end{thebibliography}\endgroup
\end{document}